\documentstyle [12pt,twoside]{article}
\skip\footins 10pt plus 4pt minus 2pt    
\font\fr=cmssdc12                          
\font\tenss=cmss12                         
\font\sevenss=cmss9                        
\font\fivess=cmss7                         
\font\sevenrm=cmr7 scaled \magstep1        
\font\fiverm=cmr5 scaled \magstep1         

\def\BIT{{\rm 1{\hbox to 0.4pt{\hss\rm l}}}}
\def\BIS{{\hbox{\sevenrm 1{\hbox to 0.3pt{\hss\sevenrm l}}}}}
\def\BISS{{\hbox{\fiverm 1{\hbox to 0.2pt{\hss\fiverm l}}}}}
\def\BI{{\mathchoice{\BIT}{\BIT}{\BIS}{\BISS}}}
\def\BGAMte{{\hbox{\tenss\char0{\hbox to 0.15em{\hss\tenss\char0}}}}}
\def\BGAMse{{\hbox{\sevenss\char0{\hbox to 0.15em{\hss\sevenss\char0}}}}}
\def\BGAMfi{{\hbox{\fivess\char0{\hbox to 0.15em{\hss\fivess\char0}}}}}

\def\BPIte{{\hbox{\tenss\char0{\hbox to 0.3em{\hss\tenss\char5}}}}}
\def\BPIse{{\hbox{\sevenss\char0{\hbox to 0.3em{\hss\sevenss\char5}}}}}
\def\BPIfi{{\hbox{\fivess\char0{\hbox to 0.3em{\hss\fivess\char5}}}}}

\def\BSIGte{{\hbox{\tenss\char6{\hbox to 0.2em{\hss\tenss\char6}}}}}
\def\BSIGse{{\hbox{\sevenss\char6{\hbox to 0.2em{\hss\sevenss\char6}}}}}
\def\BSIGfi{{\hbox{\fivess\char6{\hbox to 0.2em{\hss\fivess\char6}}}}}

\def\BBte{{\hbox{\tenss\rlap{L}\char0{\hbox to 0.27em{\hss\tenss B}}}}}
\def\BBse{{\hbox{\sevenss\rlap{L}\char0{\hbox to 0.27em{\hss\sevenss B}}}}}
\def\BBfi{{\hbox{\fivess\rlap{L}\char0{\hbox to 0.27em{\hss\fivess B}}}}}

\def\BCte{{\,\hbox{\tenss I{\hbox to 0.2em{\hss\tenss C}}}}}
\def\BCse{{\,\hbox{\sevenss I{\hbox to 0.2em{\hss\sevenss C}}}}}
\def\BCfi{{\,\hbox{\fivess I{\hbox to 0.2em{\hss\fivess C}}}}}
\def\BC{{\mathchoice{\BCte}{\BCte}{\BCse}{\BCfi}}}
\def\BDte{{\hbox{\tenss\rlap{L}\char0{\hbox to 0.35em{\hss\tenss D}}}}}
\def\BDse{{\hbox{\sevenss\rlap{L}\char0{\hbox to 0.35em{\hss\sevenss D}}}}}
\def\BDfi{{\hbox{\fivess\rlap{L}\char0{\hbox to 0.35em{\hss\fivess D}}}}}

\def\BEte{{\hbox{\tenss\rlap{L}\char0{\hbox to 0.22em{\hss\tenss E}}}}}
\def\BEse{{\hbox{\sevenss\rlap{L}\char0{\hbox to 0.22em{\hss\sevenss E}}}}}
\def\BEfi{{\hbox{\fivess\rlap{L}\char0{\hbox to 0.22em{\hss\fivess E}}}}}

\def\BFte{{\hbox{\tenss\char0{\hbox to 0.19em{\hss\tenss F}}}}}
\def\BFse{{\hbox{\sevenss\char0{\hbox to 0.19em{\hss\sevenss F}}}}}
\def\BFfi{{\hbox{\fivess\char0{\hbox to 0.19em{\hss\fivess F}}}}}

\def\BGte{{\hbox{\tenss C{\hbox to 0.15em{\hss\tenss G}}}}}
\def\BGse{{\hbox{\sevenss C{\hbox to 0.15em{\hss\sevenss G}}}}}
\def\BGfi{{\hbox{\fivess C{\hbox to 0.15em{\hss\fivess G}}}}}

\def\BHte{\,{\hbox{\tenss I{\hbox to 0.3em{\hss\tenss H}}}}}
\def\BHse{\,{\hbox{\sevenss I{\hbox to 0.3em{\hss\sevenss H}}}}}
\def\BHfi{\,{\hbox{\fivess I{\hbox to 0.3em{\hss\fivess H}}}}}

\def\BBIte{{\hbox{\tenss 1{\hbox to 0.11em{\hss\tenss 1}}}}}
\def\BBIse{{\hbox{\sevenss 1{\hbox to 0.11em{\hss\sevenss 1}}}}}
\def\BBIfi{{\hbox{\fivess 1{\hbox to 0.11em{\hss\fivess 1}}}}}

\def\BJte{{\hbox{\tenss J{\hbox to 0.13em{\hss\tenss J}}}}}
\def\BJse{{\hbox{\sevenss J{\hbox to 0.13em{\hss\sevenss J}}}}}
\def\BJfi{{\hbox{\fivess J{\hbox to 0.13em{\hss\fivess J}}}}}

\def\BKte{{\,\hbox{\tenss I{\hbox to 0.27em{\hss\tenss K}}}}}
\def\BKse{{\,\hbox{\sevenss I{\hbox to 0.27em{\hss\sevenss K}}}}}
\def\BKfi{{\,\hbox{\fivess I{\hbox to 0.27em{\hss\fivess K}}}}}

\def\BLte{{\hbox{\tenss L{\hbox to 0.15em{\hss\tenss L}}}}}
\def\BLse{{\hbox{\sevenss L{\hbox to 0.15em{\hss\sevenss L}}}}}
\def\BLfi{{\hbox{\fivess L{\hbox to 0.15em{\hss\fivess L}}}}}

\def\BMte{\,{\hbox{\tenss\char16{\hbox to 0.48em{\hss\tenss M}}}}}
\def\BMse{\,{\hbox{\sevenss\char16{\hbox to 0.48em{\hss\sevenss M}}}}}
\def\BMfi{\,{\hbox{\fivess\char16{\hbox to 0.48em{\hss\fivess M}}}}}

\def\BNte{\,{\hbox{\tenss\char16{\hbox to 0.3em{\hss\tenss N}}}}}
\def\BNse{\,{\hbox{\sevenss\char16{\hbox to 0.3em{\hss\sevenss N}}}}}
\def\BNfi{\,{\hbox{\fivess\char16{\hbox to 0.3em{\hss\fivess N}}}}}
\def\BN{{\mathchoice{\BNte}{\BNte}{\BNse}{\BNfi}}}
\def\BOte{{\,\hbox{\tenss I{\hbox to 0.3em{\hss\tenss O}}}}}
\def\BOse{{\,\hbox{\sevenss I{\hbox to 0.3em{\hss\sevenss O}}}}}
\def\BOfi{{\,\hbox{\fivess I{\hbox to 0.3em{\hss\fivess O}}}}}

\def\BPte{{\hbox{\tenss\char0{\hbox to 0.24em{\hss\tenss P}}}}}
\def\BPse{{\hbox{\sevenss\char0{\hbox to 0.24em{\hss\sevenss P}}}}}
\def\BPfi{{\hbox{\fivess\char0{\hbox to 0.24em{\hss\fivess P}}}}}

\def\BQte{{\,\hbox{\tenss I{\hbox to 0.3em{\hss\tenss Q}}}}}
\def\BQse{{\,\hbox{\sevenss I{\hbox to 0.3em{\hss\sevenss Q}}}}}
\def\BQfi{{\,\hbox{\fivess I{\hbox to 0.3em{\hss\fivess Q}}}}}
\def\BQ{{\mathchoice{\BQte}{\BQte}{\BQse}{\BQfi}}}
\def\BRte{{\hbox{\tenss\char0{\hbox to 0.25em{\hss\tenss R}}}}}
\def\BRse{{\hbox{\sevenss\char0{\hbox to 0.25em{\hss\sevenss R}}}}}
\def\BRfi{{\hbox{\fivess\char0{\hbox to 0.25em{\hss\fivess R}}}}}
\def\BR{{\mathchoice{\BRte}{\BRte}{\BRse}{\BRfi}}}
\def\BSte{{\hbox{\tenss S{\hbox to 0.13em{\hss\tenss S}}}}}
\def\BSse{{\hbox{\sevenss S{\hbox to 0.13em{\hss\sevenss S}}}}}
\def\BSfi{{\hbox{\fivess S{\hbox to 0.13em{\hss\fivess S}}}}}

\def\BTte{{\hbox{\tenss T{\hbox to 0.13em{\hss\tenss T}}}}}
\def\BTse{{\hbox{\sevenss T{\hbox to 0.13em{\hss\sevenss T}}}}}
\def\BTfi{{\hbox{\fivess T{\hbox to 0.13em{\hss\fivess T}}}}}

\def\BUte{{\,\hbox{\tenss U{\hbox to 0.15em{\hss\tenss J}}}}}
\def\BUse{{\,\hbox{\sevenss U{\hbox to 0.15em{\hss\sevenss J}}}}}
\def\BUfi{{\,\hbox{\fivess U{\hbox to 0.15em{\hss\fivess J}}}}}

\def\BVte{{\hbox{\tenss\rlap{\raise0.6ex\hbox{v}}{\hbox{\tenss V}}}}}
\def\BVse{{\hbox{\sevenss\rlap{\raise0.6ex\hbox{v}}{\hbox{\sevenss V}}}}}
\def\BVfi{{\hbox{\fivess\rlap{\raise0.6ex\hbox{v}}{\hbox{\fivess V}}}}}

\def\BWte{{\hbox{\tenss V{\hbox to 0.41em{\hss\tenss W}}}}}
\def\BWse{{\hbox{\sevenss V{\hbox to 0.41em{\hss\sevenss W}}}}}
\def\BWfi{{\hbox{\fivess V{\hbox to 0.41em{\hss\fivess W}}}}}

\def\BZte{{\hbox{\tenss Z{\hbox to 0.2em{\hss\tenss Z}}}}}
\def\BZse{{\hbox{\sevenss Z{\hbox to 0.2em{\hss\sevenss Z}}}}}
\def\BZfi{{\hbox{\fivess Z{\hbox to 0.2em{\hss\fivess Z}}}}}
\def\BZ{{\mathchoice{\BZte}{\BZte}{\BZse}{\BZfi}}}
\def\mymid{|}

\def\adj{^{\scriptscriptstyle+}}

\def\ww{\hbox{${\cal W}${\hbox to 0.2pt{\hss${\cal W}$}}}}
\def\w{{\cal W}}
\def\sw{{\cal SW}}
\def\wa#1{\w(2,{#1})}
\def\swa#1{\sw(\frac{3}{2},{#1})}
\def\n{{\cal N}}

\def\lb{\lbrack}
\def\rb{\rbrack}
\def\de{\partial}

\def\q#1{$\lb#1\rb$}

\def\sn{\smallskip\noindent}
\def\bn{\bigskip\noindent}

\def\ts{\textstyle}
\def\ds{\displaystyle}

\def\abschnitt#1#2{\par\noindent
   \hbox to \textwidth{{\bf#1}\hrulefill{\bf#2}}}
\def\longabschnitt#1#2#3{\par\noindent
   \vtop{
      \hbox to \textwidth{{\bf#1}\hrulefill{\bf#2}}
      \hbox to \textwidth{\hfill{\bf#3}}
   }
}
\def\kapitel#1#2#3{\par\noindent
   \vbox to 0.5\textheight{
      \vfill\hbox to \textwidth{{\Huge\bf#1}\hfill}
      \hbox to \textwidth{\hrulefill\vspace{1ex}}
      \hbox to \textwidth{\hfill{\Huge\bf#2}\vspace{1ex}}
      \hbox to \textwidth{\hfill{\Huge\bf#3}}
      \vfill
   }
}
\def\smallkapitel#1#2#3{\par\noindent
   \vtop{
      \hbox to \textwidth{{\large\bf#1}\hfill\vspace{-1ex}}
      \hbox to \textwidth{\hrulefill}
      \hbox to \textwidth{\hfill{\large\bf#2}\vspace{0.5ex}}
      \hbox to \textwidth{\hfill{\large\bf#3}}
   }
}

\def\smallcapitel#1{\par\noindent
   {\large\bf#1}}

\def\vac#1{\mymid{#1}\rangle}
\def\avac#1{\langle{#1}\mymid}

\def\pdssum{\phantom{{\displaystyle\sum}}}

\def\bn{\bigskip}                          
\def\strut#1#2{\rule[#1]{0cm}{#2}}         
\def\pdssum{\vphantom{{\displaystyle\sum}}}
\begin{document}
  \setcounter{page}{1}
  \normalsize\rm
\pagestyle{empty}
$\phantom{x}$\vskip 4cm\par
{\huge \begin{center}
$\w$-Algebras, New Rational Models\\
and\\
Completeness of the $c = 1$ Classification
\end{center}}\par
\vfill
\begin{center}
$\phantom{X}$\\
{\LARGE Michael Flohr}
\end{center}\par
\vfill
\begin{center}
  {\bf Abstract}
\end{center}
\begin{quotation}
  \noindent Two series of $\w$-algebras with two generators are constructed
  from chiral vertex operators of a free field representation.
  If $c = 1 - 24k$, there exists a $\wa{3k}$ algebra for $k\in\BZ_{+}/2$
  and a $\wa{8k}$ algebra for $k\in\BZ_{+}/4$. All possible lowest-weight
  representations, their characters and fusion rules are calculated proving
  that these theories are rational. It is shown, that these non-unitary
  theories complete the classification of all rational theories with
  effective central charge $c_{{\em eff}} = 1$. The results are generalized
  to the case of extended supersymmetric conformal algebras.
\end{quotation}\par\vfill\vfill
\begin{quotation}\noindent
  \begin{raggedright}\noindent
    \begin{tabular}{@{}l@{}}
      Physikalisches Institut der Universit\"at Bonn \\
      Nussallee 12                                   \\
      W-5300 Bonn 1                                  \\
      Germany                                        \\
      email:\ unp055@ibm.rhrz.uni-bonn.de            \\
    \end{tabular}
  \end{raggedright}
  \hfill
  \begin{raggedleft}
    \begin{tabular}{@{}l@{}}
      BONN-HE-92-08                 \\
      Bonn University               \\
      March 1992                    \\
      {\footnotesize ISSN-0172-8733}\\
      hep-th/9207019                \\
    \end{tabular}
  \end{raggedleft}\\
\end{quotation}\par$\phantom{x}$
\newpage
\pagestyle{plain}
  \smallkapitel{1}{Introduction}{}
  \smallcapitel{S}ince the fundamental work of Belavin, Polyakov and
Zamolodchikov \q{3} one of the most exciting problems in theoretical
physics is the classification of all possible conformal field theories
(CFT). As is well known this outstanding question plays a central
r\^{o}le in statistical physics as well as string theory and even in the
mathematics of 3-manifolds due to its conection with topological quantum
field theory \q{38}\q{31}.\par
  \sn In the last years two in some sense dual concepts of
classification where developped. One of them is the study of extended
conformal symmetry algebras, the so called $\w$-algebras as introduced
by Zamolodchikov \q{39}. In this approach one first explicitly
constructs an algebra of local chiral fields and then gets insight into the
CFT by the study of its irreducible representations. The other one
deals with abstract properties of representations of conformally
invariant operator algebras only, leaving the latter more or less
unspecified. Here one tries to construct abstract fusion algebras \q{35}.
The second approach is more restrictive since it only considers rational
conformal theories (RCFT). In this case modular invariance of partition
functions might be seen as a link between these to methods, since on the one
hand they can be constructed from the characters of the irreducible
representations of the symmetry algebras, on the other hand they assure
the existence of a unitary and symmetric $S$-matrix yielding the fusion
algebra via the famous Verlinde formula \q{37}\q{33}.\par
  \sn $\w$-algebras describe the operator product expansion (OPE) of
conformally invariant local chiral fields. The singular part of such
an OPE yields a Lie bracket structure for the Fourier modes of the
fields, the regular part an operation of forming normal ordered
products. In the following we define a $\w$-algebra as generated by
a finite set of primary fields $\phi_0,\phi_1,\ldots,\phi_n$ including the
identity, whose modes yield an associative algebra closed under derivation
$\partial$ and quasi-primary normal ordering $\n(\cdot,\cdot)$ (see
\q{5}). In addition the fields $\phi_i$ are assumed to be simple, i.e.\
not composed from others by the operation $\n$. With the conformal dimensions
$h(\phi_i) = d_i$ we denote such an algebra by $\w(2,d_1,\ldots,d_n)$, where
all structure constants are left unspecified and
$d_0 = 2$ stands for the Virasoro field instead of the identity.\par
  \sn Some of these $\w$-algebras were constructed in the last few
years by different groups implementing the conformal bootstrap
\q{22}\q{40}\q{18}\q{6}.
Recently many new examples could be investigated
using the Lie bracket approach \q{5}\q{26}, which has beside
others the great advantage of directly leading to a Lie algebra
structure thus admitting the definition of lowest-weight
representations \q{13}\q{36}.\par
  \bn In this paper we establish a new class of RCFTs using both
pictures. In the second chapter, starting from some explicitly
constructed examples of $\w$-algebras, a whole series is extracted
founded on general arguments from the theory of degenerate models.
The structure constants of these $\w$-algebras are calculated in
chapter three.
In the fourth chapter we explain the explicit calculated irreducible
representations of these examples by deriving general character formulae,
the modular invariant partition function and finally the $S$-matrix
and the fusion algebra, where details are shifted to two appendices.
This leads to a new class of RCFTs for which the whole characterizing
data is presented. It turns out that, while all these RCFTs are non-unitary,
they fit in the frame of the classification of all theories with central
charge $c = 1$. The completion of this classification towards the
non-unitary case is the subject of chapter five.
The sixth chapter gives the obvious generalization of
the models to the supersymmetric case.\par
  \bn\bn\bn
  \smallkapitel{2}{{\boldmath $\w$}-algebras and degenerate non-minimal
  models}{}
  \smallcapitel{T}his chapter mainly intends to explain the existence
of serieses of $\w(2,\delta)$-algebras at the central charge values
$c = 1 - 8\delta$ and $c = 1 - 3\delta$, which has been conjectured
in \q{5}. Here the notation means a local chiral algebra with one simple
generator in addition to the Virasoro field, whose Lie-algebra of modes
is algebraically closed under commutators and normal ordered products in the
closure of the envelopping algebra. For exact definitions and details
we refer the reader to this paper and \q{34}. We just sketch the results
obtained there, which motivated our work. The following table lists the
explicitly constructed $\w$-algebras, which are conjectured to be members
of two general series.
  $${                      
  \begin{array}{llcrlcrl}{\rm Table\ 2.1}&
    \multicolumn{7}{l}{{\scriptstyle{\rm Two\ sets\ of\ }
    \w{\rm -algebras\ to\ rational\ }c{\rm -values\ not\ contained\ in\ the\
    minimal\ series}}}\\ \hline
  \pdssum&\multicolumn{7}{l}{{\scriptstyle{\rm The\ series\ }
    \w(2,\delta){\rm\ with\ }c\, =\, 1 - 8\delta:}}\\
  &\wa{\frac{3}{2}} &\pdssum      &(c&=-11)&\pdssum      & (C_{WW}^{W})^2&=0\\
  &\wa{3}           &\pdssum      &(c&=-23)&\pdssum      & (C_{WW}^{W})^2&=0\\
  &\wa{\frac{9}{2}} &\pdssum      & c&=-35 &\pdssum      & (C_{WW}^{W})^2&=0\\
  &\wa{6}           &\pdssum      & c&=-47 &\pdssum      & (C_{WW}^{W})^2&=0\\
  &\wa{\frac{15}{2}}&\pdssum      & c&=-59 &\pdssum      & (C_{WW}^{W})^2&=0\\
  &\wa{9}           &\pdssum      & c&=-71 &\pdssum      & (C_{WW}^{W})^2&=0\\
  \pdssum&\multicolumn{7}{l}{{\scriptstyle{\rm The\ series\ }
    \w(2,\delta){\rm\ with\ }c\, =\, 1 - 3\delta:}}\\
  &\wa{2}           &\pdssum      &(c&=- 5)&\pdssum      &((C_{WW}^{W})^2&=
    \phantom{-}\frac{722}{34})\\
  &\wa{4}           &\pdssum      &(c&=-11)&\pdssum      & (C_{WW}^{W})^2&=
    -\frac{57434}{253}\\
  &\wa{6}           &\pdssum      &(c&=-17)&\pdssum      & (C_{WW}^{W})^2&=
    \phantom{-}\frac{95922436000}{43340157}\\
  &\wa{8}           &\pdssum      & c&=-23 &\pdssum      & (C_{WW}^{W})^2&=
    -\frac{127081705690919}{5974374591}\\
  &                 &\phantom{c=c}&        &\phantom{c=c}& & \\ \hline
  \end{array}}$$
The $c$-values in brackets are extensions of the obtained results to the
cases of generically, i.e.\ for all $c$-values up to finitely many exceptions,
existing $\w$-algebras. Some explicit calculations concerning the
representation theory of these generically existing algebras at the particular
$c$-values of table 2.1 may be found in \q{13} confirming the
extensions of our list. The explicit result for $\w(2,9)$ has been
obtained by \q{29}. In the last column of the table we list the squares of
the self-coupling structure constant $C_{WW}^{W}$ of the additional primary
field $W$ with dimension $\delta$. The value for the $\w(2,2)$-algebra has
been put in brackets, because this algebra exists for every central charge
and every self-coupling independently, since it always can be linearly
transformed into a copy of two commuting Virasoro-algebras. The particular
given value will be justified later. Of course, the self-coupling vanishes
necessarily for $\delta$ not even.\par
  \sn The generically existing $\w$-algebras can be identified in the
following way: $\w(2,\frac{3}{2})$ is nothing else than the
Super-Virasoro-algebra, and $\w(2,2)$ is the direct sum of two
Virasoro-algebras. The latter and $\w(2,3)$, $\w(2,4)$, and $\w(2,6)$ can be
viewed as the ``Casimir-algebras'' of the affine Kac-Moody-algebras,
or actually as the affinization of the Casimir-algebras, related
to the semi-simple Lie-algebras $A_1\oplus A_1$, $A_2$, $B_2$ or $C_2$, and
$G_2$ respective.\par
  \sn In order to explain the existence of the series of $\w$-algebras
of table 2.1
we start to review very shortly the theory of the so called Dotsenko-Fateev
degenerate models and free-field construction \q{12}.
The irreducible Virasoro lowest-weight modules of degenerate models are
not identical with the Verma modules due to null states. In this case
the operator algebra of the model is generated by primary fields which
correspond to the lowest-weight states of dimensions
  $$h_{r,s}(c) = \frac{1}{4}\left((r\alpha_- + s\alpha_+)^2 -
  (\alpha_- + \alpha_+)^2\right)\,,\eqno(2.1)$$
where we parametrized the central charge as $c = 1 - 24\alpha_0^2$ and
defined $\alpha_{\pm} = \alpha_0 \pm \sqrt{1 + \alpha_0^2}$.
In these references the Virasoro algebra and their irreducible representation
modules are constructed from Fock space representations of the Heisenberg
algebra (the free field)
  $$\lb j_{m},j_{n}\rb = n\delta_{n+m,0}\,,\eqno(2.2a)$$
built on the lowest-weight state $\vac{\alpha,\alpha_0}$ with
  $$j_{n}\vac{\alpha,\alpha_0} = 0\ \forall n<0\,,\ \
  j_{0}\vac{\alpha,\alpha_0} = \sqrt{2}\alpha\vac{\alpha,\alpha_0}\,.
  \eqno(2.2b)$$
These Fock spaces ${\cal F}_{\alpha,\alpha_0}$ obtain the structure of
Virasoro modules if the Virasoro field is defined by
  $$L(z) = \n(j,j)(z) + \sqrt{2}\alpha_{0}\de_{z}j(z)\,,\eqno(2.3a)$$
which has central charge $c = 1 - 24\alpha_0^2$. Here $\n(\cdot,\cdot)$
stands for the quasiprimary projection of a standard normal-ordered product.
The Heisenberg lowest-weight states become Virasoro lowest-weight states
with weights $h(\alpha) = \alpha^2 - 2\alpha\alpha_0$, i.e.\
  $$L_{n}\vac{\alpha,\alpha_0} = 0\ \forall n<0\,,\ \
  L_{0}\vac{\alpha,\alpha_0} = h(\alpha)\vac{\alpha,\alpha_0}\,.
  \eqno(2.3b)$$
One also can construct chiral primary conformal fields of weight $h(\alpha)$,
so called vertex operators, which map Fock spaces of different charges into
each  other, $\psi_{\alpha} : {\cal F}_{\beta,\alpha_0} \longrightarrow
{\cal F}_{\alpha+\beta,\alpha_0}$. They are given by the normal ordered
expression
  $$\psi_{\alpha}(z) =
  \exp\left(-\sum_{n>0}\sqrt{2}\alpha j_{n}\frac{z^n}{n}\right)
  \exp\left(-\sum_{n<0}\sqrt{2}\alpha j_{n}\frac{z^n}{n}\right)
  c(\alpha)z^{-\sqrt{2}\alpha j_0}\,,\eqno(2.4)$$
where $c(\alpha)$ commutes with the $j_n$, $n\neq 0$,
and maps groundstates
to groundstates. If $\alpha = \alpha_{r,s} = \frac{1}{2}(1 - r)\alpha_-
+ \frac{1}{2}(1-s)\alpha_+$, then $h(\alpha) = h_{r,s}(c)$ of (2.1).
Particular importance have the so called screening operators
$Q_{\pm}^{(n)} = \oint dz_{1}\ldots dz_{n}\psi_{\alpha_{\pm}}(z_1)\ldots
\psi_{\alpha_{\pm}}(z_n)$ with appropriate choosen integration paths.
Since $h(\alpha_{\pm}) = 1$, they have conformal dimension zero but do
change the charge of the Fock-space states to which they are applied.
With the help of these operators both, non-trivial $n$-point functions
and conformal blocks can be constructed.\par
  \sn Products of para-fields of the form (2.4) only can be well-defined for
radial ordered points, i.e.\ $\psi_{\alpha}(z_1)\psi_{\beta}(z_2)$ is
only defined for $\mymid z_1\mymid > \mymid z_2\mymid$. One gets the
other half by analytic continuation, and the chiral conformal blocks
in general become multivalued functions
$\psi_{\alpha}(z_1)\psi_{\beta}(z_2) = \varepsilon_{\alpha\beta}
\psi_{\beta}(z_2)\psi_{\alpha}(z_1)$, where $\varepsilon_{\alpha\beta}
= \exp(2\pi i\alpha\beta)$.
Two fields $\phi, \psi$ are said to be local relative to each other,
if their phase $\varepsilon_{\phi\psi} = \pm 1$ and their conformal
dimensions differ by integers or half integers. Thus, chiral local fields
necessarily must have integer or half integer weights.\par
  \sn Much work has been done to resolve the embedding structure
of the Virasoro and Fock space Verma modules to irreducible lowest-weight
modules, see e.g.\ \q{15}\q{16}. In fact, Felder showed that only on the
Fock spaces ${\cal F}_{r,s}$ with charges $\alpha_{r,s}$
screening operators can act well-defined, and can be considered as the
non-trivial coboundary operators (or so called BRST operators) on the
cohomology complex of the Fock spaces whose elements just are the irreducible
Virasoro modules ${\cal H}_{r,s}$. Indeed the screened vertex operators
  $$V_{(n'n)(m'm)}^{(l'l)}(z) = \psi_{n'n}(z)Q_{-}^{(r')}Q_{+}^{(r)} :
  {\cal F}_{m'm} \longrightarrow {\cal F}_{l'l}\,,\eqno(2.5)$$
where $l = m + n - 2r - 1\,,\ l' = m' + n' - 2r' -1$, are invariant
(up to a phase) under the action of the screening charges (BRST-invariant),
i.e.\ they map the cohomology spaces into each other. \par
  \sn In principle these facts contain all information about the CFT like
fusion rules, braid matrices or OPE structure coefficients,
see e.g.\ \q{17} for the case of symmetric theories.
Let $\w$ denote the maximal extended
symmetry algebra of the CFT (making the partition function diagonal).
If the theory is rational, the symmetry algebra is a finitely generated
$\w(d_0,d_1,\ldots,d_n)$-algebra, where $d_0 = 2$ denotes the always present
Virasoro field. If we denote with ${\cal H}^{(\lambda)}$ an irreducible
lowest-weight representation space of the $\w$-algebra, then this space
can be decomposed with respect to the Virasoro-algebra. Thus the whole
Hilbert space can be written as
  $${\cal H} = \bigoplus_{\lambda\in\Lambda}\left(
  \bigoplus_{(n'n)\in N_{\lambda}}{\cal H}_{n'n}^{(\lambda)}\otimes
  \bigoplus_{(n'n)\in N_{\lambda}}{\cal H}_{n'n}^{(\lambda)}\right)\,,
  \eqno(2.6)$$
where $\Lambda$ denotes the set of all irreducible $\w$-lowest-weight
representations and $N_{\lambda}$ the set of all irreducible Virasoro
lowest-weight representations contained in ${\cal H}^{(\lambda)}$.
Then the local fields of dimensions $\Delta = h_{n'n} +
\bar{h}_{\bar{n}'\bar{n}}$ are glued together from the screened vertex
operators,
  $$\Phi_{n'n,\bar{n}'\bar{n}}(z,\bar{z}) =
  \sum_{m',m,l',l} {\cal D}_{(n'n)(m'm)}^{(l'l)}V_{(n'n)(m'm)}^{(l'l)}(z)
  \otimes\sum_{\bar{m}',\bar{m},\bar{l}',\bar{l}}
  {\cal D}_{(\bar{n}'\bar{n})(\bar{m}'\bar{m})}^{(\bar{l}'\bar{l})}
  V_{(\bar{n}'\bar{n})(\bar{m}'\bar{m})}^{(\bar{l}'\bar{l})}(\bar{z})
  \,,\eqno(2.7)$$
where the coefficients ${\cal D}_{(n'n)(m'm)}^{(l'l)}$ are fixed up to
normalization by the requirements of locality of the OPE and crossing-symmetry
of the four-point-function. They are non-zero only, if
$\mymid n - m\mymid+1\leq l\leq m+n-1$, $l\equiv m+n-1$ mod $2$, and similarily
for $l'$. The situation simplifies drastically, if a chiral theory is
considered, since then the locality condition is extremely restrictive.
The OPE of two local chiral fields again only can contain local chiral
fields on the right hand side. Moreover, the chiral blocks have to be
local for themselves. The consequences of these strong requirements will be
worked out in the following.\par
  \bn Rational conformal field theories (RCFT) can be characterized by
the fact that they have
only a finite set of $\w$-primary fields or equivalently having
only finitely many irreducible $\w$-lowest-weight representations,
where $\w$ denotes the maximal extended symmetry algebra of the CFT.
But this in return means the following: Firstly both, the central charge as
well as the conformal dimensions of all fields, which belong to the operator
algebra, have to be rational numbers \q{2}.
And secondly, infinitely many Virasoro primaries with rational dimensions
are needed, because otherwise the characters will never be ({\em finite\/}
linear combinations of elementary) modular functions.\par
  \sn For the degenerate models this means the following. If (2.1) is
expressed in $k = \alpha_0^2$, thus $h_{r,s} = -k +
\frac{1}{4}\left((2k+1)(r^2+s^2) + 2\sqrt{k(k+1)}(r^2-s^2) - 2rs\right)$,
then we can distinguish three cases:
\begin{list}{}{}
  \item[(i)] $k,\sqrt{k(k+1)}\in\BQ$. In this case necessarily $k$ is
    of the form $\frac{(p-q)^2}{4pq}$ with $p,q\in\BN$ coprime, thus $c$
    belongs to the minimal series (including the case $c = 1$). Moreover
    $h_{r,s}\in\BQ\ \forall r,s\in\BZ$.
  \item[(ii)] $k\in\BQ\,,\ \sqrt{k(k+1)}\in\BC-\BQ$. This yields all
    rational $c$-values not contained in the minimal series. In this
    case exactly the weights $h_{r,\pm r}\in\BQ\ \forall r\in\BZ$ only.
  \item[(iii)] $k\in\BC-\BQ$. In this case neither $c$ nor the $h$-values
    are rational (the latter up to the exception $h_{1,1} = 0$).
  \end{list}\noindent
The proof of this statement is simple. First note, that the polynomials
in $r,s$ which have coefficients $k$ or $\sqrt{k(k+1)}$ respective are
linearly independent. Thus case (i) is obtained by the requirement, that
all dimensions should be rational, i.e.\ the coefficients have to be rational.
Put $\sqrt{k(k+1)} = \frac{n}{m}\in\BQ$. Solving this for $k$ and requiring
rationality yields the diophantic equation $(2n)^2 + m^2 = l^2$ with
the Pythagorian triples as their solutions. Parametrizing the coprime solutions
yields $c$ in the minimal series. Case (ii) simply comes out, if one
looks for the zeros of the polynomials. Only $(r^2 - s^2)$ admits infinitely
many solutions allowing rational $h$-values, while its coefficient is
non-rational. The last case just covers the remaining possibilities. Note,
that $k$ and $\sqrt{k^2+k}$ are algebraically independent numbers over $\BQ$
for all irrational $k$.\par
  \sn Of particular interest is case (ii). If one chooses $k$ to be
integer or half-integer one finds that all weights
$$h_{r,r} = (r^2 - 1)k\,,\ \ \ h_{r,-r} = (r^2 - 1)k + r^2\eqno(2.8)$$
are integer or half-integer. Moreover the phases $\varepsilon_{\alpha_{r,r}
\alpha_{s,s}}$ all equal $\pm 1$. If $r$ is odd, then even $k\in\BZ_{+}/4$
is possible. This shows that the ``diagonal'' set
$\{V_{(n,n)(m,m)}^{(l,l)}\, \mymid\, n,m,l\in\BZ_{+}\,,\ l\equiv n+m-1\
{\rm mod}\ 2\}$ of BRST-invariant screened vertex operators is a local
set, i.e.\ all operators are local to each other since the phases
appearing by reordering the screening charges cancel as long as the
number of reordered $Q_+$ charges equals the number of reordered $Q_-$
charges, $Q_{-}^{(r)}Q_{+}^{(r)}\psi_{\alpha_{n,n}}(z) =
e^{4\pi ir\alpha_{n,n}\alpha_0}\psi_{\alpha_{n,n}}(z)Q_{-}^{(r)}Q_{+}^{(r)}$
and $4\alpha_{n,n}\alpha_0 = 4(1-n)\alpha_0^2\in\BZ$ for
$\alpha_0^2 = k \in \BZ_{+}/4$. From now on we consider this special case
of ``diagonal'' operators, i.e.\ $n' = \pm n$ in $\alpha_{n,n'}$.
In the following we use the shorter notation
$V_{n,m}^{l}(z) \equiv V_{(n,n)(m,m)}^{(l,l)}(z)$ for the ``digonal''
BRST-invariant screened vertex operators (2.5) and
${\cal D}_{n,m}^{l} \equiv {\cal D}_{(n,n)(m,m)}^{(l,l)}$ for their
coefficients in the chiral blocks $W^{(n)} \equiv\Phi_{n,n,1,1}$ according
to (2.7).\par
  \sn Indeed, the chiral blocks must be glued together from the
operators of the local set above. Otherwise they cannot represent
chiral local fields. Locality also restricts the fusion rules for the
chiral algebra, since chiral local operators map the spaces
into each other such that
the lowest weights differ by integers or half-integers. Thus, the OPE
of two of these fields acting on a lowest-weight module again only
yields local fields acting on lowest-weight modules. This implies that
the set of local chiral blocks,
  $$\left\{\left.W^{(n)}(z) = \sum_{m\in\BZ_{+}}
  \sum_{{l\in\BZ/(n+m)\BZ \atop l+n+m\,\equiv\,1\,{\rm mod}\,2}}
  {\cal D}_{n,m}^{l}V_{n,m}^{l}(z)\, \right\mymid\,
  n\in\BZ_{+}\right\}\eqno(2.9a)$$
is a closed algebra with fusion rules
  $$\left[W^{(n)}\right] \times \left[W^{(m)}\right] =
  \sum_{{\mymid n-m\mymid+1\,\leq\,l\,\leq\,m+n+1 \atop
  l+m+n+1\,\equiv\,1\,{\rm mod}\,2}}N_{n,m}^{l}
  \left[W^{(l)}\right]\,,\eqno(2.9b)$$
where the fusion numbers $N_{n,m}^{l}$ are non negative integers.
Moreover, the subset with $n$ odd is a closed subalgebra, and will
be called the odd sector of the algebra in the following.\par
  \sn It is important to notice that only (half-) integer $k$ (or
quarter-integer for the odd sector subalgebra) will lead to non-trivial
RCFTs. One could think to take for $k$ other rational numbers than
these and to look for the subset of chiral vertex operators that are
still local. But in this case it can happen that e.g.\ two operators
$W^{(n)}$ and $W^{(m)}$ having (half-) integer dimensions are local with
respect to each other, while one of them, say $W^{(n)}$, being not local to
itself. Then it might happen, that the conformal family of such an operator
$W^{(n)}$ contributes to the
right hand side of the OPE of the other local field with itself. In this
case, the simple field does not appear on the right hand side but its normal
ordered
products, e.g.\ $\n(W^{(n)},W^{(n)})$ which has to be understood as the chiral
projection of the normal ordered product of the left-right symmetric
field $W^{(n)}(z)\oplus W^{(n)}(\bar{z})$ with itself. As a consequence,
no algebraically closed local algebra larger than the Virasoro algebra
can be defined. Indeed, we will see later that the values $k\in\BZ_{+}/4$
are the only possibilities to obtain non-minimal RCFTs from Dotsenko-Fateev
models.\par
  \bn It remains to show that these local algebras are indeed $\w$-algebras,
i.e.\ completely generated by the normal ordered products of (derivatives of)
finitely many simple primary fields. From the fusion rules we learn that
applying a field twice to a lowest-weight state will lead us to other
lowest-weight states corresponding to other local primary fields. Indeed,
the complete local system can be generated by application of $W^{(2)}$, the
odd sector by using $W^{(3)}$. On the other hand we can consider the
commutator of modes of two operators and use the truncation of the terms on the
right hand side by the conformal dimension rather than by the label of
the Fock space charge. Writing the right hand side symbolically in
conformal families we find e.g.\
  $$\left[W^{(2)}_{m},W^{(2)}_{n}\right] =
  \left[W^{(1)}_{m+n}\right] + \left[W^{(3)}_{m+n}\right]
  \,,\eqno(2.10)$$
where $W^{(1)}$ of course is the identity operator. For the
dimensions we have $h_{3,3} = 8k > 2h_{2,2} - 1 = 2(3k) - 1$ with
$k = (1-c)/24$. Thus no field of the conformal family
$\left[W^{(3)}\right]$ will appear in the commutator of $W^{(2)}$
with itself showing that the modes of the latter field together with
the Virasoro modes generate a Lie-algebra structure which closes in
(the closure of) its envelopping algebra, actually they generate a
$\w(2,3k)$ algebra. The same argumentation applies to the odd sector algebras,
where one can eliminate $W^{(5)}$ from the right hand side of the commutator
of $W^{(3)}$ with itself, yielding a $\w(2,8k)$ algebra.\par
  \sn The associativity of the OPE is equivalent with the fact that
the Jacobi identities are fullfilled. As was pointed out in \q{5}
only the identities involving three simple fields have to be checked
and there only the coefficients for the primary fields appearing on
the right hand side. This leaves us with one non-trivial condition in
our case. If the simple field has dimension $\delta$, then fields up to
dimension $3\delta - 2$ will appear on the right hand side of the
Jacobi identity. Comparing again with the fusion rules we see for
the odd sector algebras that
$3h_{3,3} - 2 = 24k - 2 < h_{5,5} = 24k$ showing that no further
primary field can contribute to the identity. In the other case
we have $3h_{2,2} - 2 = 9k - 2 \ge h_{3,3} = 8k$ for $k \ge 2$
indicating that the
field $W^{(3)}$ could contribute to the identity. But its coefficient
must be zero because the self-coupling of $W^{(2)}$ vanishes due to
the fusion rules. In fact, vanishing self-coupling means that no field
can appear on the right hand side, whose mode expansion has monomials
involving more than one mode of $W^{(2)}$. But the composite primary
field $W^{(3)}$, which is nothing than the primary projection of
$\n(W^{(2)},\de^{2k}W^{(2)})$, will have terms quadratic in $W^{(2)}$ in
its mode expansion.
Thus, if one weakens the assumptions in the
definition of $\w$-algebras such that the generators need not be
simple, then also a $\w(2,3k,8k)$ can be constructed, where $W^{(3)}$
is given as above.
Note, that if $h_{2,2}$ is half-integer, so is $k$ such that the
statement above remains valid, since $2k$ is odd as it must be.
This explains the existence of the $\w$-algebras listed in table 2.1.
  \par\bn A $\w$-algebra is completely determined by the set of
dimensions of the generators and a consistent choice of all free
parameters. The dimension of the additional primary field and the
central charge $c$ are already fixed in our cases, the only free parameter
left is the self-coupling structure constant of the primary field. In
the case of the $\wa{3k}$-algebras it must vanish by symmetry, but for
the case of $\wa{8k}$-algebras one might be interested in a formula
expressing it by the only real input, the number $k$, which also
paramterizes the central charge $c = 1 - 24k$ and the dimension $\delta = 8k$.
This and the still remaining determination of the
coefficients in the chiral conformal blocks is done in the next chapter.
  \bn\bn\bn
  \smallkapitel{3}{Structure Constants}{}
  \smallcapitel{W}e now come to the calculation of the structure constants.
In particular, we show that the self-coupling structure constants $C_{WW}^{W}$
of the $\w(2,8k)$-algebras can be derived from the structure constants of
the Dotsenko-Fateev models. This will make our proof rigorous that these
algebras can be represented by a free-field construction. In the following
we use the notation as in \q{17}.\par
  \sn As was pointed out in \q{5}\q{34},
the commutators in a $\w$-algebra
are fixed by SU(1,1)-invariance up to some structure constants. Furthermore,
all structure constants for quasiprimary fields can be reduced by
SU(1,1)-invariance to expressions involving only the central charge $c$
and the structure constants connecting three simple primary fields.
In the case of a $\w(2,\delta)$-algebra with only one additional primary
field $W$, there is only one up to now free structure constant beside
the central charge, the self-coupling constant $C_{WW}^{W}$. Its square
is usually fixed by the validity of the Jacobi identity involving three
times the field $W$, which in this case is the only identity one has to
check.\par
  \sn Since every chiral local theory can be tensored with itself to yield
a symmetric theory, we learn from (2.7) that the coefficients of the
symmetric Dotsenko-Fateev models, as given in \q{17} and
denoted there as $D_{NM}^{L}$, have to equal the squares of our
${\cal D}_{n,m}^{l}$ with $N = (n,n), M = (m,m)$ and $L = (l,l)$ (up to
normalization).
Felder, Fr\"ohlich and Keller
obtained the following result from the calculation
of the braid matrices of the BRST-invariant vertex operators, which are
proportional to the (quantum) $6j$-symbols of the quantum group
U$_q$(SU(2)), and the crossing symmetry of the latter:
  $$\begin{array}{rcl}
    {\ds\left({\cal D}_{n,m}^{l}\right)^2} &=&
    {\ds c\cdot\frac{h_{l,l}}{h_{n,n}h_{m,m}}D_{NM}^{L}\ =\
    c\cdot\frac{h_{l,l}}{h_{n,n}h_{m,m}}
    \frac{N_{LL}^{(1,1)}}{N_{NN}^{(1,1)}N_{MM}^{(1,1)}}
    \Delta_{n,m}^{l}(x)\Delta_{n,m}^{l}(x')}\,,
    \strut{-3ex}{3ex}\\
    {\ds\Delta_{n,m}^{l}(x)} &=& {\ds (-1)^{\frac{1}{2}(n+m-l-1)}
    \left(\frac{[n]_x[m]_x[l]_x}{[1]_x}\right)^{\frac{1}{2}}}
    \strut{-3ex}{5ex}\\
    &\times&{\ds\prod_{j=(l+n-m+1)/2}^{n-1}[j]_x
    \prod_{j=(m+n-l+1)/2}^{n-1}[j]_x
    \prod_{j=(l+m-n+1)/2}^{(l+m+n-1)/2}\frac{1}{[j]_x}}\,,
    \strut{0ex}{4ex}
  \end{array}\eqno(3.1)$$
where the brackets are given by $[j]_x = x^{j/2} - x^{-j/2}$ with
$x = \exp(2\pi i\alpha_+^2)$ and $x' = \exp(2\pi i\alpha_-^2)$.
A prefactor $c\cdot h_{l,l}h_{n,n}^{-1}h_{m,m}^{-1}$
has been included to take care of our
normalization of the two-point functions used in \q{5}\q{34},
which is defined for chiral, simple, primary fields to take the value
  $$\avac{0}W^{(n)}_{-h_{n,n}}W^{(m)}_{h_{m,m}}\vac{0} =
  \frac{c}{h_{n,n}}\delta_{n,m}\,.\eqno(3.2)$$
The general normalization constants $N_{(n'n)(m'm)}^{(l'l)} =
\avac{h_{l'l}}V_{(n'n)(m'm)}^{(l'l)}(1)\vac{h_{m'm}}$
have been expressed by Felder in terms of Dotsenko-Fateev integrals
\q{12} and are given here for completeness:
  $$\begin{array}{rcl}
    {\ds N_{(n'n)(m'm)}^{(l'l)}} &=&
    {\ds (-1)^{\frac{1}{2}((2n'-1)r+(2n-1)r')}\alpha_{+}^{4rr'}
    \prod_{j'=1}^{r'}\frac{[m'-j']_{x'}[j']_{x'}}{[1]_{x'}}
    \prod_{j=1}^{r}\frac{[m-j]_x[j]_x}{[1]_x}}
    \strut{-3ex}{3ex}\\
    &\times&{\ds\prod_{j'=1}^{r'}\frac{\Gamma(j'\alpha_-^2)
    \Gamma(m+(j'-m')\alpha_-^2)}{\Gamma(\alpha_-^2)
    \Gamma(m+n-2r+(r'-m'-n'+j')\alpha_-^2)}}
    \strut{-3ex}{5ex}\\
    &\times&{\ds\prod_{j=1}^{r}\frac{\Gamma(j\alpha_+^2-r')
    \Gamma(m'-r'+(j-m)\alpha_+^2)}{\Gamma(\alpha_+^2)
    \Gamma(m'-r'+n'+(r-m-n+j)\alpha_+^2)}}\,,
    \strut{0ex}{4ex}
  \end{array}\eqno(3.3)$$
where $l = n + m - 2r - 1$ and similar for $l'$.
The structure constants of the OPE or equivalently of the Lie-algebra
of the Fourier modes of the chiral local fields are then given by
  $$C_{n,m}^{l} =
  {\cal D}_{n,m}^{l}N_{NM}^{L}\,.\eqno(3.4)$$
Thus, in the case of our $\w(2,8k)$-algebras we find that the square
of the self-coupling of the additional simple primary field
$W = W^{(3)}$ with dimension $\delta = 8k$ reads
  $$\left(C_{WW}^{W}\right)^2 = \frac{c}{\delta}
  \left(D_{(3,3)(3,3)}^{(3,3)}N_{(3,3)(3,3)}^{(3,3)}\right)^2\,.
  \eqno(3.5)$$
Note that only the square of the structure constant can be determined
by (3.1). Since $N_{(2,2)(2,2)}^{(2,2)} = 0$ due to the fusion rules,
the self-coupling of $W^{(2)}$ vanishes as expected.
If one expresses the brackets as $[j]_x = 2i\sin(j\pi\alpha_+^2)$
and $[j]_{x'} = 2i\sin(j\pi\alpha_-^2)$, and reduces the Gamma-functions
to terms of the form $\Gamma(z)\Gamma(1-z) = \frac{\pi}{\sin(\pi z)}$,
one finally arrives at the closed expressions
  $$\left(C_{WW}^{W}\right)^2 = \left\{
  \begin{array}{l}
    {\ds\frac{(1-24k)}{8k}
        \frac{\prod_{j=1}^{8k}\left(j^2-64(k^2+k)\right)^2
              \prod_{j=1}^{2k}\left(j^2-4(k^2+k)\right)^3}
             {\prod_{j=1}^{6k}\left(j^2-36(k^2+k)\right)
              \prod_{j=1}^{4k}\left(j^2-16(k^2+k)\right)^4}}
    \strut{-5ex}{5ex}\\
    \phantom{--------}{\rm if}\ k\in\BZ_{+}/2\,,
    \strut{-2ex}{3ex}\\
    {\ds\frac{(1-24k)}{8k}
        \frac{\prod_{j=1}^{8k}\left(j^2-64(k^2+k)\right)^2
              \prod_{j=1}^{2k+1/2}\left((j-\frac{1}{2})^2-4(k^2+k)\right)^3}
             {\prod_{j=1}^{6k+1/2}\left((j-\frac{1}{2})^2-36(k^2+k)\right)
              \prod_{j=1}^{4k}\left(j^2-16(k^2+k)\right)^4}
        \frac{3}{4(k^2+k)}}
    \strut{-5ex}{5ex}\\
    \phantom{--------}{\rm if}\ k\in\BZ_{+}+\frac{1}{4}\,.
  \end{array}\right.\eqno(3.6)$$
This result agrees with the examples of table 2.1.
The value for the $\w(2,2)$ given there has been obtained from (3.6).
While this algebra exists for every central charge and self-coupling
independently, it is related to a Dotsenko-Fateev model for $c = -5$
only for this particular value of the structure constant. Other values
could be obtained, if the Virasoro field (2.3) would be deformed.\par
  \sn One remark is necessary here. The screened vertex operators
$V_{(n'n)(m'm)}^{(l'l)}(z)$ in (2.5) carry a representation of the braid group,
namely the braid matrices given by \q{17} and defined by the
relation
  $$V_{KL}^{J}(z)V_{NM}^{L}(w) = \sum_{L'}R(J,K,N,M)_{LL'}
    V_{NL'}^{J}(w)V_{KM}^{L'}(z)\eqno(3.7)$$
valid for $\mymid w\mymid > \mymid z\mymid$. The ordering of the integration
variables of the screening charges and the choice of their contours are of
great importance for the calculation of the braid matrices. In our case, where
$n'=n$ etc., one always has the same number of positive and negative
screening charges. Thus,
one can introduce similar vertex operators ${\cal V}_{n,m}^{l} =
\psi_{n,n}(\widetilde{Q})^{(r)} : {\cal F}_{m,m} \rightarrow {\cal F}_{n,n}$
with $\widetilde{Q} = Q_-Q_+$, i.e.\ with
a rearrangement of the ordering of the screening charges and a change of
their contours: Applied to a vertex operator located at $z$, the new
screening operator is given by $\widetilde{Q} =
\oint_{z}du\int_{z}^{u}du'\psi_{\alpha_-}(u')\psi_{\alpha_+}(u)$,
where the inner integration over $u'$ follows the same contour
as the outer one over $u$ which encircles zero and starts at $z$. Then
these operators look like fields of a thermal theory, i.e.\ a theory
with $N = (1,n)$, $M = (1,m)$ etc., but with a double integration for
every effective screening with $\widetilde Q$.
Since the braid matrices are almost factorized in a
left and right thermal part,
their components connecting only ``diagonal''
operators are independent from the non-diagonal ones.
Moreover, with
the modified operators, the (thermal) braid matrices $r(j,k,n,m\mymid x)$
in \q{17} simplify drastically
by taking the limit $x \rightarrow 1$, since the effective phase of
moving contours of the effective screening charge $\widetilde{Q}$ is
$\alpha_+^2 + \alpha_-^2 = 2k + 1 \in \BZ_{+}/2$ for our particular models.
In this limit $[m]_x/[n]_x \rightarrow m/n$ thus leaving us with simple
rational numbers for the matrix elements and the
${\cal D}_{n,m}^{l}$ coefficients. On the other hand the
behaviour of the analytic continuation of the normalization integrals
also changes, actually simplifies, if they are defined in the modified
vertex operators, since the latter have trivial monodromy properties.
Of course both effects cancel out in formula (3.4) leaving the structure
constants unchanged as it should be. But this remark shows the special
r\^ole of the values $k \in \BZ_{+}/4$ of the background charge: For
these values the modified screened vertex operators form a very simple
representation of the braid group.\par
  \sn In the next chapter the explicit calculation of all allowed lowest-weight
representations together with a modular invariant partition function
completes the description of these new CFTs and proves that they are indeed
rational.
  \bn\bn\bn
  \smallkapitel{4}{Representation Theory}{}
  \smallcapitel{I}n this chapter we discuss the representation theory
of the $\w$-algebras established in the last chapter. The answering of
this question yields the complete field content of $\w$-primary fields
of the theory analogous to the case of the Virasoro algebra.
Starting from the
character of the vacuum representation of the $\w$-algebra we find all
lowest-weight-representations by considering the behaviour of this
character under modular transformations.\par
  \sn The case of the bosonic $\w$-algebras in the $1 - 8d$ series
is treated in detail, for the other series only a brief discussion and
the results are given.\par
  \sn In the previous chapter we have shown that for $c \in \BQ$ but $c$
not an element of the minimal series, only the degenerate conformal
families with weights $h_{n,n} = (n^2 - 1)\frac{1 - c}{24}$ and
$h_{n,-n} = (n^2 - 1)\frac{1 - c}{24} + n^2$ have rational conformal
weights. In particular for $c = 1 - 24k\,,\ k\in\BN/2$ all these fields
have integer or half-integer dimension, which is necessary for building
chiral local symmetry algebras.\par
  \sn From now on let $c = 1 - 24k$ with $k\in\BN/2$ fixed.
Since the $\w$-algebra which is infinitely generated by the primary
fields belonging to the weights $h_{n,n}$ for $c = 1 - 24k$ contains the
$\w(2,3k)$-algebra as finitely generated subalgebra, all primary fields
with higher spin have to be composite. This follows using the isomorphism
between the Hilbert space of the vacuum representation of the
$\w(2,3k)$-algebra, generated by the modes of the two simple fields, and the
space of all quasiprimary fields, which can be generated by normal ordered
products of (derivatives of) the simple fields. As deduced in the last
chapter, the other primary fields
then appear as primary projections of normal ordered products according
to the fusion rules. For example the field $W = W^{(2)}$ obeys the
fusion rule
  $$[W^{(2)}]\times[W^{(2)}] = [\BI] + [W^{(3)}]\,,\eqno(4.1)$$
where the weights are $h_{2,2} = 3k$ and $h_{3,3} = 8k$. This means
that the conformal family $[W^{(3)}]$ does not occur in the commutator
(or equivalently the singular part of the OPE)
of $W$ with itself, while the primary projection of $\n(W,\de^{2k}W)$
is proportional to $W^{(3)}$.\par
  \sn Thus, remembering that all these primary fields belong to degenerate
conformal families created by singular vectors, the $\w$-algebra character
of the vacuum representation is given by summing up all the Virasoro characters
of the lowest-weight representations $\vac{h_{n,n}}$. Following the work
of Feigin and Fuks \q{15}, every Virasoro lowest-weight module at
level $h_{n,n}$ has exactly one null vector at level
$h_{n,-n} = h_{n,n} + n^2$. Therefore, if $\chi_{h_{n,n}}(\tau)$
denotes the character of such a lowest-weight representation of the
Virasoro algebra, it is given by
  $$\begin{array}{lcl}
  \chi_{h_{n,n}}(\tau) &=& {\ds\frac{q^{(1-c)/24}}{\eta(\tau)}
                           \left(q^{h_{n,n}} - q^{h_{n,-n}}\right)}
                           \strut{-3ex}{3ex}\\
                       &=& {\ds\frac{1}{\eta(\tau)}
                           \left(q^{n^{2}k} - q^{n^{2}(k + 1)}\right)}
                           \strut{0ex}{4ex}\,,
  \end{array}\eqno(4.2)$$
where $q = e^{2\pi i\tau}$ as usual, $\tau$ being the modular parameter
of the torus, and the Dedekind $\eta$-function is
$\eta(\tau) = q^{1/24}\prod_{n=1}^{\infty}(1-q^n)$.
This implies that the $\w$-algebra character can be written in the form
  $$\begin{array}{lcl}
  \chi^{\w}_{0}(\tau)  &=& {\ds\sum_{n\in\BZ_{+}}\chi_{h_{n,n}}(\tau)}
                           \strut{-3ex}{3ex}\\
                       &=& {\ds\frac{1}{2\eta(\tau)}\sum_{n\in\BZ}
                           \left(q^{n^{2}k} - q^{n^{2}(k + 1)}\right)}
                           \strut{-3ex}{7ex}\\
                       &=& {\ds\frac{1}{2\eta(\tau)}\left(
                           \Theta_{0,k}(\tau) - \Theta_{0,k+1}(\tau)\right)}
                           \strut{0ex}{4ex}\,,
  \end{array}\eqno(4.3)$$
where we have introduced the elliptic functions (modular functions of weight
one-half)
  $$\Theta_{\lambda,k}(\tau) = \sum_{n\in\BZ}q^{(2kn + \lambda)^2/4k}\,,\ \ \
  \lambda \in \BZ_{+}/2\,,\ k \in \BN/2\,.\eqno(4.4a)$$
We call $\lambda$ the index and $k$ the modulus of the function.
Surprisingly, we can express our $\w$-algebra character by functions with
well known properties under modular transformations, actually they will
form a finite dimensional representation space of the modular group.
Indeed we will show that these $\w$-algebras belong to a RCFT. Note
that in contrast to the known cases (e.g.\ minimal models, WZW models)
elliptic functions of different moduli are involved.\par
  \sn Let us additionally introduce the alternating elliptic functions
  $$\widetilde{\Theta}_{\lambda,k}(\tau) =
  \sum_{n\in\BZ}(-1)^nq^{(2kn + \lambda)^2/4k}\,,\ \ \
  \lambda \in \BZ_{+}/2\,,\ k \in \BN/2\,.\eqno(4.4b)$$
Then the modular properties are given by
  $$\begin{array}{lcl}
  \Theta_{\lambda,k}({\ts-\frac{1}{\tau}}) &=& {\ds
    {\ts\sqrt{\frac{-i\tau}{2k}}}\,\sum_{\lambda'=0}^{2k-1}
    e^{i\pi\frac{\lambda\lambda'}{k}}\left\{
    \begin{array}{ll}
      \Theta_{\lambda',k}(\tau)         & \mbox{if $\lambda\in\BZ$}\\
      \widetilde{\Theta}_{\lambda',k}(\tau) &
        \mbox{if $\lambda\in\BZ + \frac{1}{2}$}
    \end{array}\right.}\strut{-3ex}{3ex}\,,\\
  \widetilde{\Theta}_{\lambda,k}({\ts-\frac{1}{\tau}}) &=& {\ds
    {\ts\sqrt{\frac{-i\tau}{2k}}}\,\sum_{\lambda'=0}^{2k-1}
    e^{i\pi\frac{\lambda(\lambda' + \frac{1}{2})}{k}}\left\{
    \begin{array}{ll}
      \Theta_{\lambda' + \frac{1}{2},k}(\tau) & \mbox{if $\lambda\in\BZ$}\\
      \widetilde{\Theta}_{\lambda' + \frac{1}{2},k}(\tau) &
        \mbox{if $\lambda\in\BZ + \frac{1}{2}$}
    \end{array}\right.}\strut{-3ex}{7ex}\,,\\
  \Theta_{\lambda,k}({\ts\tau + 1}) &=& {\ds
    e^{i\pi\frac{\lambda^2}{2k}}\left\{
    \begin{array}{ll}
      \Theta_{\lambda,k}(\tau)         & \mbox{if $\lambda - k\in\BZ$}\\
      \widetilde{\Theta}_{\lambda,k}(\tau) &
        \mbox{if $\lambda - k\in\BZ + \frac{1}{2}$}
    \end{array}\right.}\strut{-3ex}{7ex}\,,\\
  \widetilde{\Theta}_{\lambda,k}({\ts\tau + 1}) &=& {\ds
    e^{i\pi\frac{\lambda^2}{2k}}\left\{
    \begin{array}{ll}
      \widetilde{\Theta}_{\lambda,k}(\tau) & \mbox{if $\lambda - k\in\BZ$}\\
      \Theta_{\lambda,k}(\tau)         &
        \mbox{if $\lambda - k\in\BZ + \frac{1}{2}$}
    \end{array}\right.}\strut{-3ex}{7ex}\,,\\
    \eta({\ts-\frac{1}{\tau}}) &=& {\ds\sqrt{-i\tau}\,\eta(\tau)}
      \strut{-2ex}{6ex}\,,\\
    \eta({\ts\tau + 1})        &=& {\ds e^{\pi i/12}\,\eta(\tau)}
      \strut{0ex}{3ex}\,.
  \end{array}\eqno(4.5)$$
Thus, the functions $\Lambda_{\lambda,k}(\tau) = \Theta_{\lambda,k}(\tau)
/\eta(\tau)$ are modular forms of weight zero to some $\Gamma(N) \subset
{\rm PSL}(2,\BZ)$, e.g.\ $N$ is the lowest common multiple of $4k$ and
$24$, if $\lambda - k\in\BZ$. As was argued by Kiritsis \q{27} the
Serre-Stark theorem assures the completeness of the set
$\{\Lambda_{\lambda,k} \mymid k\in\BN/2,\ 0\leq\lambda\in\BZ/2\leq k\}$
as a generating set for 1-singular modular forms as characters of RCFTs
with $c_{{\em eff}} \leq 1$ are supposed to be (the {\em effective} central
charge $c_{{\em eff}}$ will be defined later).\par
  \sn We now have to identify the characters of the other representations,
which are labelled by the pairs $(\lambda,k),\ 0\leq\lambda\leq k$ and
$(\lambda,k+1),\ 0\leq\lambda\leq k+1$. Obviously one can write
  $$\begin{array}{lcl}
  \Lambda_{\lambda,k}(\tau)   &=& {\ds\frac{q^{(1-c)/24}}{\eta(\tau)}
     \sum_{n\in\BZ}q^{h_{n+\frac{\lambda}{2k},n+\frac{\lambda}{2k}}}}
     \strut{-3ex}{3ex}\,,\\
  \Lambda_{\lambda,k+1}(\tau) &=& {\ds\frac{q^{(1-c)/24}}{\eta(\tau)}
     \sum_{n\in\BZ}q^{h_{n+\frac{\lambda}{2k+2},-n-\frac{\lambda}{2k+2}}}}
     \strut{0ex}{4ex}\,.
  \end{array}\eqno(4.6)$$
Here we have expressed the contributing lowest-weight values by formula (2.1)
for numerating the degenerate weights by pairs of integers, but used
rational non-integral numbers for the labeling except for the case
$\lambda = 0$, which corresponds to the
vacuum representation. Therefore, for $\lambda\neq 0$ there will be no
null states in the corresponding Virasoro lowest-weight modules.
Consequentely, we identify the characters for the $\w$-algebra
lowest-weight representations
$\vac{h_{\frac{\lambda}{2k},\frac{\lambda}{2k}}}$ and
$\vac{h_{\frac{\lambda}{2k+2},-\frac{\lambda}{2k+2}}}$
just to equal (see Table 4.1)
$\chi_{\lambda}^{\w}(\tau)  = \Lambda_{\lambda,k}(\tau),\ 1\leq\lambda<k$,
$\chi_{-\lambda}^{\w}(\tau) = \Lambda_{\lambda,k+1}(\tau),\ 1\leq\lambda<k+1$
and
$\chi_{k}^{\w}(\tau)    = \frac{1}{2}\Lambda_{k,k}(\tau)$,
$\chi_{-k-1}^{\w}(\tau) = \frac{1}{2}\Lambda_{k+1,k+1}(\tau)$,
where the factor $1/2$ in the last two characters removes an unphysical
double-counting of all states. Of course we still need one character
which can be identified to be
$\chi_{k+1}^{\w}(\tau) = \frac{1}{2}(\Lambda_{0,k}(\tau)
+ \Lambda_{0,k+1}(\tau))$. This can be seen as follows: It is clear
that there must be one other linear combination of $\Lambda_{0,k}$ and
$\Lambda_{0,k+1}$ except the one for the vacuum character. Other
$\Lambda$-functions cannot be combined, because their $q$-powers will never
differ by integers. Since the character is supposed not to involve
degenerate multiplicities of the corresponding representation, its
$q$-expansion has to start with leading coefficient $1$. This restricts
the possibilities to the ansatz
$\chi_{k+1}^{\w} = \mu\Lambda_{0,k} \pm (1 - \mu)\Lambda_{0,k+1}$.
The requirement of integer coefficients further restricts $\mu$ to the
set $\{0,\frac{1}{2},1\}$. But the solutions $\mu = 0$ or $\mu = 1$
again yield an unphysical double-counting of all states with weights
greater than zero,
$\mu = \frac{1}{2}$ is left as the only possibility. This means physically
that the vacuum representation lives on the invariant subspaces
left after dividing out the modules generated by the null states
(see \q{16}), while the other representation (which has the lowest-weight
$h_{{\em min}} = -k < 0 = h_{{\em vac}}$ demonstrating the non-unitarity
of the theory) lives on the direct sum of the
whole Verma modules together with the modules
generated by the singular vectors.\par
  \sn If one now expresses the modular properties in the basis of the
characters found so far, one finds a $S$-matrix which is neither symmetric
nor unitary. This comes from a hidden degeneracy of the representations
on the lowest-weight states $\vac{h_{\frac{1}{2},\pm\frac{1}{2}}}$, as
can be seen from the modular invariant partition function
  $$\begin{array}{lcl}
  Z(\tau,\bar{\tau}) &=& {\ds
    \frac{1}{2}\sum_{\lambda=0}^{2k-1}
    \left\mymid\Lambda_{\lambda,k}\right\mymid^2 +
    \frac{1}{2}\sum_{\lambda'=0}^{2k}
    \left\mymid\Lambda_{\lambda',k+1}\right\mymid^2}
    \strut{-3ex}{3ex}\\
                     &=& {\ds
    \sum_{\lambda=0}^{k-1}
    \left\mymid\chi_{\lambda}^{\w}\right\mymid^2 +
    \sum_{\lambda'=0}^{k}
    \left\mymid\chi_{-\lambda'}^{\w}\right\mymid^2 +
    2\left\mymid\chi_{k}^{\w}\right\mymid^2 +
    2\left\mymid\chi_{-k-1}^{\w}\right\mymid^2}
    \strut{0ex}{4ex}
  \end{array}\eqno(4.7)$$
which directly shows up the multiplicities. The reason for these degeneracies
lies in the extended Cartan subalgebra. Indeed, in the explicitly computed
examples \q{5}\q{13} of $\w$-algebras we found that exactly for these
representations the $W_{0}$-eigenvalue is non-zero. Actually, since the
selfcoupling of the $W$-field with itself is zero in all cases, only
the value $w^2$ given by $W_{0}W_{0}\vac{h,w} = w^2\vac{h,w}$ can be
computed by expressing $\n(W,W)$ in terms of (normal ordered products of)
the Virasoro field yielding $w^2$ as a function in $h$ and $c$ if its
zero mode is applied to the lowest-weight state. Of course, $w^2\neq 0$
will give two representations $\vac{h,\pm\sqrt{w^2}}$.\par
  \sn In order to calculate the fusion algebra, one now either has to
use the Verlinde formula \q{37}\q{33} in the modified form
  $$N_{ij}^{k} = n_i n_j \sum_m \frac{S_{im}S_{jm}{S\adj}_{km}}{S_{0m}}
  \eqno(4.8)$$
for a generalized diagonal modular invariant partition function
  $$Z(\tau,\bar{\tau}) = \sum_{m} n_m \left\mymid\chi_m\right\mymid^2
  \,,\ \ \ n_m\in\BZ_{+}\,,\eqno(4.9)$$
or one has to extend the $S$-matrix and the number of characters removing
the degeneracies (see \q{32}). This second method means in our case,
where we have two representations with multiplicities $2$, that there is
a doubling of the characters $\chi_{k,+}^{\w} = \chi_{k,-}^{\w} \equiv
\chi_{k}^{\w}$
and $\chi_{-k-1,+}^{\w} = \chi_{-k-1,-}^{\w} \equiv \chi_{-k-1}^{\w}$. Thus,
one has to extend the $S$-matrix by two rows and columns. The requirements
symmetry, unitarity and $S$ being at most of order $4$, i.e.\
  $$S = S^t\,,\ \ \ SS\adj = \BI\,,\ \ \ S^2 = C\,,$$
where $C$ is the conjugation matrix, already fix $S$ up to three free
constants which can uniquely determined from the $N_{ij}^{k}$, which
should be non-negative integers. This leads to the $S$- and $T$-matrix
given in appendix A. The fusion algebra and a calculation of the three
free parameters of the $S$-matrix are given in appendix B.
Here and in the sequel we use the following notation for the characters
and representations
  $$\begin{array}{ll|c||rcl|ll}
  \multicolumn{2}{l}{{\rm Table\ 4.1\ }}
    & \multicolumn{6}{l}{{\scriptstyle{\rm
    representations\ and\ their\ characters\ for\ the\ bosonic\ }
    \w(2,3k){\rm -algebras}}} \\ \hline
  \phantom{xxx} & \multicolumn{6}{l}{\phantom{-}} & \phantom{xxx} \\
  & h \strut{-1ex}{1ex}                                & w^2
    & \chi_{\lambda}^{\w}                              &
    &
    & {\scriptstyle{\rm remark}}                       & \\ \cline{2-7}
  & h_{1,1} \strut{-1ex}{4ex}                          & 0
    & \chi_{0}^{\w}                                    & =
    & \frac{1}{2}(\Lambda_{0,k} - \Lambda_{0,k+1})
    & {\scriptstyle{\rm vacuum\ rep.}}                 & \\
  & h_{\frac{1}{2k},\frac{1}{2k}} \strut{-1ex}{1ex}    & 0
    & \chi_{1}^{\w}                                    & =
    & \Lambda_{1,k}
    &                                                  & \\
  & \vdots                                             & \vdots
    &                                                  & \vdots
    &
    &                                                  & \\
  & h_{\frac{k-1}{2k},\frac{k-1}{2k}} \strut{-1ex}{1ex}& 0
    & \chi_{k-1}^{\w}                                  & =
    & \Lambda_{k-1,k}
    &                                                  & \\
  & h_{\frac{1}{2},\frac{1}{2}} \strut{-1ex}{1ex}      & \neq 0
    & \chi_{k,+}^{\w} = \chi_{k,-}^{\w}                & =
    & \frac{1}{2}\Lambda_{k,k}
    & {\scriptstyle{\rm degenerate\ rep.}}             & \\
  & h_{0,0} \strut{-1ex}{1ex}                          & 0
    & \chi_{k+1}^{\w}                                  & =
    & \frac{1}{2}(\Lambda_{0,k} + \Lambda_{0,k+1})
    & {\scriptstyle{\rm rep.\ to\ }h_{{\em min}}}      & \\
  & h_{\frac{1}{2k+2},-\frac{1}{2k+2}}\strut{-1ex}{1ex}& 0
    & \chi_{-1}^{\w}                                   & =
    & \Lambda_{1,k+1}
    &                                                  & \\
  & \vdots \strut{-1ex}{1ex}                           & \vdots
    &                                                  & \vdots
    &
    &                                                  & \\
  & h_{\frac{k}{2k+2},-\frac{k}{2k+2}}\strut{-1ex}{1ex}& 0
    & \chi_{-k}^{\w}                                   & =
    & \Lambda_{k,k+1}
    &                                                  & \\
  & h_{\frac{1}{2},-\frac{1}{2}} \strut{-2ex}{2ex}     & \neq 0
    & \chi_{-k-1,+}^{\w} = \chi_{-k-1,-}^{\w}          & =
    & \frac{1}{2}\Lambda_{k+1,k+1}
    & {\scriptstyle{\rm degenerate\ rep.}}             & \\ \cline{2-7}
  \multicolumn{8}{l}{\phantom{-}} \\ \hline\end{array}$$
which completely explains the representations of the $\w(2,3k)$-algebras
for $c = 1 - 24k$, $k\in\BN$ found in \q{13}. Note the change in the
labelling of the representations due to the multiplicities. These
characters diagonalize the modular invariant partition functions,
while it is maximal non-diagonal expressed in terms of Virasoro
characters. Thus these $\w(2,3k)$-algebras are very good examples
that extending the symmetry algebra does make the modular invariant
partition function more diagonal and can yield new RCFTs not related
to minimal models or any coset construction. Moreover, these non-diagonal
partition functions are not contained in the $ADE$-classification
of Cappelli, Itzykson and Zuber \q{8} and probably not related
to any other non-diagonal invariant coming from affine Lie algebras.\par
  \sn From (4.5) we learn, that for $k,\lambda\in\BZ$ the functions
$\Lambda_{\lambda+\frac{1}{2},k}$ and $\Lambda_{\lambda+\frac{1}{2},k+1}$
built a space invariant under $T^2$ and $S^2$. These functions are the
characters of the irreducible lowest-weight representations of the so called
twisted bosonic $\w(2,3k)$-algebra which is obtained by using half-integer
fourier-modes, hence introducing antiperiodic boundary conditions.
In the twisted sector of the bosonic $\w(2,3k)$-algebras no linear combinations
of these functions are necessary nor possible. Consequently the characters of
the lowest-weight representations
$\vac{h_{\frac{2\lambda+1}{4k},\frac{2\lambda+1}{4k}}}$ are simply
$\chi^{\w}_{\lambda+\frac{1}{2}}(\tau) =
\Lambda_{\lambda+\frac{1}{2},k}(\tau)$, $0\leq\lambda<k$,
and the ones of the lowest-weight representations
$\vac{h_{\frac{2\lambda+1}{4k+4},-\frac{2\lambda+1}{4k+4}}}$ read
$\chi^{\w}_{-\lambda-\frac{1}{2}}(\tau) =
\Lambda_{\lambda+\frac{1}{2},k+1}(\tau)$, $0\leq\lambda<k+1$.\par
  \sn Indeed, for some $\w$-algebras as examples these representations
could be found by explicit calculations in \q{13}.\par
  \bn Let us emphasize one point here. We show these theories
to be RCFTs by constructing the $S$-Matrix and calculating the
fusion rules. If one introduces the effective value of the central charge
  $$c_{{\em eff}} = c - 24h_{{\em min}}\,,\eqno(4.10)$$
one can compare non-unitary theories with unitary ones. Actually, the
central charge $c$ is nothing else than the mean expectation value
of the Casimir effect contribution of the free energy due to the boundary
conditions. As is well known, from the modular invariant partition function
  $$Z(\tau,\bar{\tau}=\tau) =
  {\rm Tr}\, e^{2\pi i\tau(L_0 - \frac{c}{24})}
  e^{2\pi i\bar{\tau}(\bar{L}_0 - \frac{c}{24})}
  \eqno(4.11)$$
one easily derives the following expression for the central charge
in dependency from the energy $L_0 + \bar{L}_0$
(the latter being defined up to an arbitrary additive constant)
  $$c = 12\frac{\tau\langle L_0 + \bar{L}_0\rangle_{\tau}
     -\frac{1}{\tau}\langle L_0 + \bar{L}_0\rangle_{-\frac{1}{\tau}}}
     {\tau - \frac{1}{\tau}}\eqno(4.12)$$
which simplifies for the fixed point of the $S$-transformation, $\tau = i$.
Now we obtain with $\Delta_n = h_n + \bar{h}_n$
  $$c = 12\frac{\sum_{n}\Delta_{n}\exp(-2\pi\Delta_{n})}
  {\sum_{n}\exp(-2\pi\Delta_{n})}\,,\eqno(4.13)$$
where the sum extends over the weights of all states, both the
conformal dimensions of the primary fields and the weights of all their
descendents.\par
  \sn This formula shows that positive exponents will appear in
non-unitary theories, since the state of lowest energy is not identical
with the vacuum, thus violating the conservation of probability. Therfore
it does make sense to redefine the energy by subtracting the energy of
the ground state,
$(L_0 + \bar{L}_0)_{\em eff} = L_0 + \bar{L}_0 - 12\Delta_{\em min}$,
which in return forces to redefine the central charge
  $$c_{\em eff} = 12\frac{\sum_{n}(\Delta_{n}-\Delta_{\em min})\exp(-2\pi
  (\Delta_{n}-\Delta_{\em min}))}
  {\sum_{n}\exp(-2\pi(\Delta_{n}-\Delta_{\em min}))}
  \,,\eqno(4.14)$$
such that the characters will keep unchanged. This effective central charge
measures the mean expectation value of the Casimir effect contribution
of the free energy relatively to
the state of lowest energy. This is a physical observable which does
conserve probability and can be used for both unitary and non-unitary
theories as well. Consequentely $c_{\em eff} > 0$.
Usually one considers symmetric theories with $\bar{h} = h$ corresponding
to diagonal modular invariant partition functions. In this case
$\Delta_{\em min} = 2h_{\em min}$.\par
  \sn In particular our
theories have $c_{{\em eff}} = 1$ and thus complete the classification
of all rational theories with $c = 1$, given in \q{27}\q{20}\q{9},
including the non-unitary case, since these theories exactly represent
the only possible additional case found in \q{27} but rejected
there due to the unnecessary restrictive assumption,
the state of lowest energy
would always be the vacuum. Note that most of the theorems used in
the references given above are valid for the non-unitary case as well,
one only has to distinguish carefully between the vacuum representation
and the minimal representation.
Quantum dimensions for example have to
be defined with respect to the conformal $\w$-family to the Virasoro
lowest weight $h_{{\em min}}$
rather than to the identity family, if the theory is non-unitary.
The next chapter is devoted to
the proof of the statement concerning the $c_{{\em eff}} = 1$ models.\par
  \bn Now we will briefly discuss the case of the fermionic
$\w(2,3k)$-algebras.
Here $c = 1 - 24\frac{k}{2}$, again $k \in \BN$. The vacuum representation
belongs to the Neveu-Schwarz sector. For this sector we find that all the
characters can be expressed in terms of the functions
$\Lambda_{\lambda,\frac{k}{2}}(\tau)$ and
$\Lambda_{\lambda,\frac{k}{2}+1}(\tau)$ with $\lambda\in\BZ$. Conversely,
for the Ramond sector we have $\lambda\in\BZ+\frac{1}{2}$. Using the
modular properties given in (4.5), it is easy to see that the Neveu-Schwarz
sector is invariant under the transformations $S$ and $T^2$, while the
Ramond sector is under $T$ and $ST^2S$, and that the transformation $TST$
intertwines the two sectors. In particular, we obtain for the NS-sector
  $$\Lambda_{\lambda,\frac{k}{2}}({\ts -\frac{1}{\tau}}) =
  \frac{1}{\sqrt{k}}\sum_{\lambda'=0}^{k-1}e^{2i\pi\frac{\lambda\lambda'}{k}}
  \Lambda_{\lambda',\frac{k}{2}}(\tau)\,.\eqno(4.15a)$$
Therefore we just can take the $S$-matrix for the bosonic case, as given
in appendix A, change every occurence of $k$ to $\frac{k}{2}$ and
remove both the $(k,+)$-th and $(k,-)$-th row and column as well as the
$(-k-1,+)$-th and $(-k-1,-)$-th ones. Note that degenerate
representations do not appear in the NS-sector.\par
  \sn For the R-sector the situation is not as simple, because the
appropriate transformation matrix is $\widetilde{S} = ST^2S$. Using again
(4.5) and eliminating one summation by a Gauss-sum results in
  $$\Lambda_{\lambda+\frac{1}{2},\frac{k}{2}}({\ts -\frac{\tau}{1-2\tau}}) =
  \frac{1}{\sqrt{k}}\sum_{\lambda'=0}^{k-1}
  e^{-i\pi\frac{(\lambda+\lambda'+1)^2}{2k}}
  \frac{1 + (-i)^{2(\lambda+\lambda'+1) + k}}{1 + (-i)}
  \Lambda_{\lambda'+\frac{1}{2},\frac{k}{2}}(\tau)\,.\eqno(4.15b)$$
Now one might go through the same procedure as for the bosonic case
with this matrix $\widetilde{S}$ and remove the degeneracies of two of the
representations, but we will not go into further detail here, since it is
not clear whether $ST^2S$ can be used instead of $S$ for calculating the
fusion rules via the Verlinde formula, nor what should replace the identity
representation.\par
  \bn Here we also briefly mention the $\w(2,8k)$-algebra series, which
exists for all $k\in\BN/4$. As has been explained in chapter 2 these
algebras just represent the odd sector of the $\w(2,3k)$-algebras (note
that there is no algebra built from the odd and even sector together
for $k\in\BN+\frac{1}{4}$ due to violation of locality). The vacuum
character for the odd sector reads
  $$\begin{array}{lcl}
  \chi_{0,{\em odd}}^{\w}(\tau) & = & {\ds\frac{q^{(1-c)/24}}{\eta(\tau)}
    \sum_{{r\in\BN\atop r\equiv 1\, {\rm mod}\, 2}}\left(
    q^{h_{r,r}} - q^{h_{r,-r}}\right)}\strut{-4ex}{4ex}\\
  & = & {\ds\frac{1}{2\eta(\tau)}\left(\Theta_{4k,4k}(\tau) -
    \Theta_{4k+4,4k+4}(\tau)\right)}\,.\end{array}\eqno(4.16)$$
The modular transformations involve the other functions
$\Theta_{\lambda+4k+4\varepsilon,4k+4\varepsilon}$ where $\varepsilon
= 0$ or $1$. For $0\leq\lambda\leq 4k+4\varepsilon$ this is a complete
set of linear independent Theta-functions. Rewriting the summations
as sums over odd integers only, the label has to be multiplied by
two, i.e.\ the lowest-weights are parametrized as
$h_{\frac{2\lambda}{2(4k+4\varepsilon)},
(-1)^{\varepsilon}\frac{2\lambda}{2(4k+4\varepsilon)}}$, since
  $$\begin{array}{lcl} {\ds
  \frac{1}{\eta(\tau)}\Theta_{\lambda+4k+4\varepsilon,4k+4\varepsilon}(\tau)}
  & = & {\ds\frac{1}{\eta(\tau)}
  \sum_{n\in\BZ}q^{\frac{\lb 2(4k+4\varepsilon)n
  + (\lambda + 4k + 4\varepsilon)\rb^2}{4(4k + 4\varepsilon)}}}
  \strut{-3ex}{3ex}\\
  & = & {\ds\frac{1}{\eta(\tau)}
  \sum_{n\in\BZ}q^{\lb (2n+1) + \frac{2\lambda}{2(4k+4\varepsilon)}\rb^2
  (k + \varepsilon)}}\strut{-3ex}{7ex}\\
  & = & {\ds\frac{q^{(1-c)/24}}{\eta(\tau)}
  \sum_{r\in(2\BZ+1)+\frac{2\lambda}{2(4k+4\varepsilon)}}
  q^{(r^2-1)k + r^2\varepsilon}}\strut{0ex}{4ex}\,.
  \end{array}\eqno(4.17)$$
Some of the characters are linear combinatios of elliptic functions to
different moduli like in the bosonic case discussed in detail above.
In particular we have
  $$\begin{array}{ll|c||rcl|ll}
  \multicolumn{2}{l}{{\rm Table\ 4.2\ }}
    & \multicolumn{6}{l}{{\scriptstyle{\rm
    representations\ and\ their\ characters\ for\ the\ bosonic\ }
    \w(2,8k){\rm -algebras}}} \\ \hline
  \phantom{xxx} & \multicolumn{6}{l}{\phantom{-}} & \phantom{xxx} \\
  & h \strut{-1ex}{1ex}                                & w
    & \chi_{\lambda,{\em odd}}^{\w}                    &
    &
    & {\scriptstyle{\rm remark}}                       & \\ \cline{2-7}
  & h_{1,1} \strut{-1ex}{4ex}                          & 0
    & \chi_{0}^{\w}                                    & =
    & \frac{1}{2}(\Lambda_{4k,4k} - \Lambda_{4k,4k+4})
    & {\scriptstyle{\rm vacuum\ rep.}}                 & \\
  & h_{\frac{1}{4k},\frac{1}{4k}} \strut{-1ex}{1ex}    & \neq 0
    & \chi_{1}^{\w}                                    & =
    & \Lambda_{1+4k,4k}
    &                                                  & \\
  & \vdots                                             & \vdots
    &                                                  & \vdots
    &
    &                                                  & \\
  & h_{\frac{4k-1}{4k},\frac{4k-1}{4k}} \strut{-1ex}{1ex}& \neq 0
    & \chi_{4k-1}^{\w}                                 & =
    & \Lambda_{8k-1,4k} = \Lambda_{-1,4k}
    &                                                  & \\
  & h_{2,2} \strut{-1ex}{1ex}                          & \neq 0
    & \chi_{4k}^{\w}                                   & =
    & \frac{1}{2}(\Lambda_{0,4k} - \Lambda_{0,4k+4})
    & {\scriptstyle{\rm rep.\ on\ }\vac{W^{(2)}}}      & \\
  & h_{0,0} \strut{-1ex}{1ex}                          & \neq 0
    & \chi_{4k+1}^{\w}                                 & =
    & \frac{1}{2}(\Lambda_{0,4k} + \Lambda_{0,4k+4})
    & {\scriptstyle{\rm rep.\ to\ }h_{{\em min}}}      & \\
  & h_{1,1} \strut{-1ex}{1ex}                          & \neq 0
    & \chi_{4k+2}^{\w}                                 & =
    & \frac{1}{2}(\Lambda_{4k,4k} + \Lambda_{4k,4k+4})
    & {\scriptstyle\vac{h=0,w\neq 0}{\rm\ rep.}}       & \\
  & h_{\frac{1}{4k+4},-\frac{1}{4k+4}}\strut{-1ex}{1ex}& \neq 0
    & \chi_{-1}^{\w}                                   & =
    & \Lambda_{1+4k+4,4k+4}
    &                                                  & \\
  & \vdots \strut{-1ex}{1ex}                           & \vdots
    &                                                  & \vdots
    &
    &                                                  & \\
  & h_{\frac{4k+3}{4k+4},-\frac{4k+3}{4k+4}}\strut{-2ex}{2ex}& \neq 0
    & \chi_{-4k-3}^{\w}                                & =
    & \Lambda_{8k+7,4k+4} = \Lambda_{-1,4k+4}
    &                                                  & \\ \cline{2-7}
  \multicolumn{8}{l}{\phantom{-}} \\ \hline\end{array}$$
Note that there is no need for degenerate representations. Indeed, all
representations have multiplicity one. The reason is that the
$W_0$ eigenvalue is uniquely expressable as a function in $h,c$, and the
non-vanishing self-coupling $C_{WW}^{W}$. Again one considers the
zero-modes of $\n(W,\de^{n}W)$, $n = 0,2$, applied to the vacuum and
solves the resulting quadratic equations for $w$. These patterns also
explain the existence of two representations with $h = 0$, only one of
them being the vacuum representation, in \q{13}.\par
  \sn Finally we note that the general addition law of the
Jacobi-Riemann $\Theta$-functions to moduli containing a square factor,
  $$\sum_{\nu=0}^{n-1}\Lambda_{n\lambda+\nu,n^{2}k}(\tau) =
  \Lambda_{\lambda,k}(\tau)\,,\ \ \ k\in\BZ_+/2\,,\eqno(4.18)$$
shows that the whole $\w(2,3k)$-algebra can be regarded as built from
their odd sector algebra $\w(2,8k)$. In fact, the representation on a
lowest weight $\vac{h(\lambda)}$ of the whole algebra is obtained by
applying the odd sector to both, the lowest-weight state and the state
$(\Phi_{2,2})_{0}\vac{h(\lambda)}$ since for the characters we have the
relation $\Lambda_{2\lambda,4k}(\tau) + \Lambda_{2\lambda+1,4k}(\tau) =
\Lambda_{\lambda,k}(\tau)$.
  \bn\bn\bn
  \smallkapitel{5}{Classification of {\boldmath $c_{{\bf\em eff}} = 1$}
    theories}{}
  \smallcapitel{N}ow we come to the completion of the classification of
all RCFTs with $c = 1$ by considering non-unitary models. As has been
explained in the last chapter, one has to use $c_{{\em eff}} =
c - 24h_{{\em min}}$ instead of the central charge for non-unitary
theories. In the works of \q{27}\q{9}\q{20}
all unitary models with
$c = 1$ have been identified. But the proofs of the statements of these
works are not affected by the assumption of unitarity as long as one
keeps in mind that the vacuum representation is not necessarily the one
with the minimal lowest weight. Since there are strong indications that
modular forms to non-congruence subgroups of the modular group will have
infinite denominators in their Fourier expansions, we assume that
non-congruence subgroups do not lead to RCFTs.
Thus, there is only one additional
candidate for a $c=1$ model (see \q{27}). It has the partition function
  $$Z = \frac{1}{2}\left(Z(R_1) + Z(R_2)\right)\,,\eqno(5.1)$$
where $Z(R)$ denotes the partition function of an U(1)-theory of mappings
of the unit sphere $S^1 \rightarrow S^1$ with radius $R$, given by
  $$Z(R) = \frac{1}{\eta(\tau)\eta(\bar{\tau})}
  \sum_{m,n\in\BZ}\left(q^{\frac{1}{8R^2}(n+2mR^2)^2}
            \bar{q}^{\frac{1}{8R^2}(n-2mR^2)^2}\right)\,.\eqno(5.2)$$
If $2R^2 \in \BN$, then this partition function can be expressed in the
elliptic functions given by equations (4.4), namely
  $$Z(R) = \frac{1}{\eta(\tau)\eta(\bar{\tau})}
  \sum_{1\leq n\leq 4R^2}\left\mymid\Theta_{n,2R^2}\right\mymid^2\,,
  \eqno(5.3)$$
where the Theta-functions satisfy $\Theta_{n,2R^2} = \Theta_{-n,2R^2}
= \Theta_{n+4R^2,2R^2}$ and $\Theta_{2R^2,2R^2}$ has
only even integer coefficients, considered as power series in $q$.
If now $2R^2 = \frac{P}{Q} \in \BQ$ with $P,Q$ coprime, then we can write
  $$Z(R) = \frac{1}{\eta(\tau)\eta(\bar{\tau})}
  \sum_{n\,{\rm mod}\,2PQ}\Theta_{n,PQ}(\tau)\Theta_{n',PQ}(\bar{\tau})\,,
  \eqno(5.4)$$
with $n'$ given by $n' = QN + PM$ mod $2PQ$, if $n = QN - PM$ mod $2PQ$
for some integers $N,M$. Note, that the integer case $Q = 1$ is correctly
obtained from the general one.\par
  \sn It is now necessary for obtaining a RCFT from the partition
function (5.1) to have $2R_i^2 = \frac{P_i}{Q_i}$, $i = 1,2$.
This yields the following two possibilities:
  $$\begin{array}{rcl}
    Z &=& {\ds (\eta\bar{\eta})^{-1}\left(
      \sum_{n=1}^{P_{1}Q_{1}-1}\Theta_{n,P_{1}Q_{1}}
                         \bar{\Theta}_{n',P_{1}Q_{1}}
    + \sum_{m=1}^{P_{2}Q_{2}-1}\Theta_{m,P_{2}Q_{2}}
                         \bar{\Theta}_{m',P_{2}Q_{2}}
    \right.}\strut{-3ex}{3ex}\\
      &+& {\ds\left\mymid\frac{\Theta_{0,P_{1}Q_{1}}+\Theta_{0,P_{2}Q_{2}}}{2}
      \right\mymid^2
    + \left\mymid\frac{\Theta_{0,P_{1}Q_{1}}-\Theta_{0,P_{2}Q_{2}}}{2}
      \right\mymid^2
    }\strut{-3ex}{7ex}\\
    &+& \left\{\begin{array}{l} {\ds\left.
      2\left\mymid\frac{\Theta_{P_{1}Q_{1},P_{1}Q_{1}}}{2}\right\mymid^2
    + 2\left\mymid\frac{\Theta_{P_{2}Q_{2},P_{2}Q_{2}}}{2}\right\mymid^2\right)
    }\strut{-3ex}{7ex}\\ {\ds\left.
      \left\mymid\frac{\Theta_{P_{1}Q_{1},P_{1}Q_{1}}
                     + \Theta_{P_{2}Q_{2},P_{2}Q_{2}}}{2}\right\mymid^2
    + \left\mymid\frac{\Theta_{P_{1}Q_{1},P_{1}Q_{1}}
                     - \Theta_{P_{2}Q_{2},P_{2}Q_{2}}}{2}\right\mymid^2\right)
    \,.}\strut{0ex}{3ex}
    \end{array}\right.\strut{0ex}{7ex}\end{array}\eqno(5.5)$$
It is clear that only the both linear combinations with a minus sign
could be Virasoro vacuum characters, since the latter must have the form
$q^{h-c/24}(1 - q + \ldots)\bar{q}^{\bar{h}-c/24}(1 - \bar{q} + \ldots)$.
Thus we can distinguish two cases:
$$\begin{array}{rccl}
  {\rm (i)\ }    & {\ds\frac{\Theta_{0,P_{1}Q_{1}} - \Theta_{0,P_{2}Q_{2}}}{2}}
                 &=
                 & {\ds q^{P_{1}Q_{1}} - q^{P_{2}Q_{2}} + q^{4P_{1}Q_{1}}
                          + \ldots
                 }\,,\strut{-3ex}{3ex}\\
  {\rm (ii)\ }   & {\ds\frac{\Theta_{P_{1}Q_{1},P_{1}Q_{1}}
                           - \Theta_{P_{2}Q_{2},P_{2}Q_{2}}}{2}}
                 &=
                 &{\ds q^{\frac{P_{1}Q_{1}}{4}} - q^{\frac{P_{2}Q_{2}}{4}}
                           + q^{\frac{9P_{1}Q_{1}}{4}} + \ldots
                 }\,.\strut{0ex}{3ex}
  \end{array}\eqno(5.6)$$
In case (i) we obtain the condition $P_{2}Q_{2} = P_{1}Q_{1} + 1$ and hence
$c = 1 - 24P_{1}Q_{1}$, in case (ii) one has to satisfy
$P_{2}Q_{2} = P_{1}Q_{1} + 4$ and hence $c = 1 - 6P_{1}Q_{1}$.
These are exactly our series of $c$-values for the bosonic $\w(2,3k)$-algebras
with $k = P_{1}Q_{1}$ and for the odd-sector algebras $\w(2,8k)$ with
$k = P_{1}Q_{1}/4$. In fact, we have seen that under special assumptions
on the radii $R_i$, $i=1,2$, Virasoro characters can be found in the
partition function (5.1). The modular invariance of the latter and their
well known decomposition shows that the theory is rational. Even more
the extended symmetry algebra for this theory is known and can be
identified with a certain $\w$-algebra.\par
  \sn Let us remark that there can be a lot of decompositions
of the modulus in (5.4) into two coprime numbers $P,Q$.
These different decompositions
yield the non trivial automorphisms of the fusion numbers or equivalently
the set of theories, which have related partition functions (5.5) with
$n'$ and $m'$ given as described above. For details see appendix B.\par
  \sn Finally we conjecture that the set of these theories lies dense
in the set of all theories with partition function (5.1) to arbitrary
radii $R_1,R_2 \in \BR_+$. This conjecture is equivalent to the following
problem: For every positive real numbers $R_1,R_2$ and every $\varepsilon > 0$
find pairs of coprime integers $P_1,Q_1$ and $P_2,Q_2$ such that
$\mymid 2R_i^2 - \frac{P_i}{Q_i}\mymid < \varepsilon$, $i = 1,2$, and
$P_{2}Q_{2} - P_{1}Q_{1} = 1$ holds.
  \bn\bn\bn
  \smallkapitel{6}{Generalization to the supersymmetric case}{}
  \smallcapitel{T}heme of this chapter is a brief sketch of the rather
straightforward generalization of the new RCFTs to the supersymmetric
case. To fix the notation we set $c = \frac{3}{2} - 24k = \frac{3}{2}(
1 - 16k) = \frac{3}{2}\widehat{c}$ and again consider the case $k\in\BN/4$,
which will lead to theories with $c_{{\em eff}} = \frac{3}{2}$.
With $\alpha_{\pm} = \sqrt{k} \pm \sqrt{k+\frac{1}{2}}$ we have the
lowest-weight levels
  $$h_{r,s}(c) = \frac{1}{4}(r\alpha_+ + s\alpha_-)^2
  + \frac{1}{16}(\widehat{c} - 1) + \frac{1}{32}(1 - (-1)^{r-s})\,,
  \eqno(6.1)$$
where for the NS-sector $r-s\equiv 0$ mod $2$ and $\equiv 1$ mod $2$ for the
R-sector respective. As for the ordinary case, we first list the up to now
known resluts, which have been obtained by explicit calculations
\q{24}\q{30}\q{23}\q{19}\q{4}\q{14}.
Here we used the common notation where the smaller
dimension of the super-partners are denoted only, namely $\swa{\delta} =
\wa{\frac{3}{2},\delta,\delta+\frac{1}{2}}$. This is a supersymmetric
conformal algebra extended by one additional covariant supersymmetric
field $\Phi(z,\theta) = \phi(z) + \theta\psi(z)$ with dimension
$(\delta,\delta+\frac{1}{2})$, where $\theta$ denotes a Grassman variable.
In analogy to (2.9) we denote the super conformal blocks by $\Phi^{(n)}$.
The $\swa{3k}$-algebras are then formed by the field $\Phi^{(2)}$, the
$\swa{8k}$-algebras by $\Phi^{(3)}$, where we use similar arguments as
those of the second chapter.
  $${                      
  \begin{array}{llcrlcrl}{\rm Table\ 6.1}&
    \multicolumn{7}{l}{{\scriptstyle{\rm Two\ sets\ of\ }
    \sw{\rm -algebras\ to\ rational\ }c{\rm -values\ not\ contained\
    in\ the\ supersymmetric\ minimal\ series}}}\\ \hline
  \pdssum&\multicolumn{7}{l}{{\scriptstyle{\rm The\ series\ }
    \swa{\delta}{\rm\ with\ }c\, =\, \frac{3}{2} - 8\delta:}}\\
  &\swa{\frac{3}{2}} &\pdssum      &(c&=-\frac{21}{2})&\pdssum      &
    ((C_{\Phi\Phi}^{\Phi})^2&=0)\\
  &\swa{3}           &\pdssum      & c&=-\frac{45}{2} &\pdssum      &
    (C_{\Phi\Phi}^{\Phi})^2&=0\\
  &\swa{\frac{9}{2}} &\pdssum      & c&=-\frac{69}{2} &\pdssum      &
    (C_{\Phi\Phi}^{\Phi})^2&=0\\
  &\swa{6}           &\pdssum      & c&=-\frac{93}{2} &\pdssum      &
    (C_{\Phi\Phi}^{\Phi})^2&=0\\
  \pdssum&\multicolumn{7}{l}{{\scriptstyle{\rm The\ series\ }
    \swa{\delta}{\rm\ with\ }c\, =\, \frac{3}{2} - 3\delta:}}\\
  &\swa{2}           &\pdssum      &(c&=-\frac{ 9}{2})&\pdssum      &
    (C_{\Phi\Phi}^{\Phi})^2&=\phantom{-}\frac{242}{13}\\
  &\swa{4}           &\pdssum      & c&=-\frac{21}{2} &\pdssum      &
    (C_{\Phi\Phi}^{\Phi})^2&=-\frac{508369}{2499}\\
  &\swa{6}           &\pdssum      & c&=-\frac{34}{2} &\pdssum      &
    (C_{\Phi\Phi}^{\Phi})^2&=\phantom{-}\frac{6309688448}{3137409}\\
  &                  &\phantom{c=c}&                  &\phantom{c=c}&
    & \\ \hline
  \end{array}}$$
The $c$-value of some of the algebras has been put in brackets:
The $\swa{\frac{3}{2}}$-algebra does exist generically and for independently
choosen self-coupling. It is the supersymmetric analogon to the $\w(2,2)$,
and thus nothing else than a direct sum of two supersymmetric Virasoro
algebras. But only for vanishing self-coupling it is related to a
supersymmetric free field construction due to the fusion rules of the
latter. The $\swa{2}$-algebra exists for generic central charge. This
seems natural since the classical counterpart of this algebra is the
symmetry algebra of the Super-Toda theory corrsponding to the Super-Lie-algebra
{\fr osp}$(3\mymid 2)$. These, and $\swa{\frac{1}{2}}$ (Super-Kac-Moody
algebra) and $\swa{1}$ ($N=2$ Super-Virasoro algebra)
are the only known super-$\w$-algebras with two generators,
which exist for generically choosen central charge.\par
  \sn Motivated by the analogy of these series to the conformal case
we consider again the ``diagonal'' fields
with weights $h_{r,r} = (r^2-1)k$ and $h_{r,-r} = (r^2-1)k + \frac{1}{2}r^2$
in the NS-sector. In the R-sector these weights have to be shifted,
$h = h_{r,\pm r} + \frac{1}{16}$.
Checking the locality conditions for chiral theories yields exactly the
same pattern as in the non-supersymmetric case. $r=2, k\in\BN/2$ gives
the analogon of the $\w(2,3k)$-algebras, the
${\cal S}\w(\frac{3}{2},3k)$-algebras,
and $r=2, k\in\BN/4$ the analogon to the so-called odd sector subalgebras,
${\cal S}\w(\frac{3}{2},8k)$.\par
  \sn Let us first consider the $\swa{3k}$-theories. Again we start from
the vacuum representation and get all other irreducible lowest-weight
representations by modular transformations. The vacuum character is given by
  $$\begin{array}{lcl}
  \chi_{0}^{\cal{SW}}(\tau) &=& {\ds\prod_{n\in\BN}
  \frac{1+q^{n-\frac{1}{2}}}{1-q^n}q^{\frac{\widehat{c}}{16}}\sum_{r\in\BZ}
  \left(q^{h_{r,r}} - q^{h_{r,-r}}\right)}\strut{-3ex}{3ex}\\
  &=&{\ds\frac{\eta(\frac{\tau+1}{2})}{\eta^2(\tau)}e^{-\frac{\pi i}{24}}
  \left(\Theta_{0,k}(\tau) - \Theta_{0,k+\frac{1}{2}}(\tau)\right)}
  \strut{0ex}{4ex}\,.\end{array}\eqno(6.2)$$
The NS-sector turns out to be again invariant under $S$ and $T^2$, using
the $\Theta_{\lambda,k+\frac{\varepsilon}{2}}$-functions with $\lambda\in\BZ$.
The R-sector is a little bit more complicated. Here the combinatorial
prefactor making the character a modular form of weight zero is
$\prod\frac{1+q^n}{1-q^n} = \eta(2\tau)/\eta^2(\tau)$. From (4.5) we
learn, that invariance under $T$ enforces
$\lambda-k+\frac{\varepsilon}{2}\in\BZ$. Thus, the index $\lambda$ is
integer or half-integer, if the modulus $k$ is integer or half-integer
respective. Then the R-sector is invariant under $T$ and $ST^2S$.\par
  \sn Note that we only consider the characters without fermion number
counting $(-)^F$
since these are enough to classify the possible representations. Of course,
in the modular invariant partition function the characters of the
$\widetilde{{\rm NS}}$-sector,
given by ${\rm tr}_{\vac{h}}(-)^{F}q^{L_{0}-c/24}$, have to be added
to get invariance under the full modular group. But
the latter are easy to obtain from the characters of the ordinary NS-sector
without fermion number by applying the $T$-transformation to them,
$\chi^{\sw}_{\lambda,\widetilde{{\rm NS}}}(\tau) =
\chi^{\sw}_{\lambda,{\rm NS}}(\tau + 1)$.
They involve the $\widetilde{\Theta}$-functions (4.4b) instead of the
the ordinary $\Theta$-functions (4.4a) and get the prefactor
$\prod\frac{1-q^{n-\frac{1}{2}}}{1 - q^n} = \exp(-\frac{2}{16}\pi i)
\eta(\frac{1}{2}\tau)\eta^{-2}(\tau)$. Thus, they are essentially
given by the functions
  $$\Lambda^{\widetilde{{\rm NS}}}_{\lambda,k} =
  \frac{\eta(\frac{\tau}{2})}{\eta^2(\tau)}e^{-\frac{2\pi i}{16}}
  \widetilde{\Theta}_{\lambda,k}(\tau)\,,\eqno(6.3)$$
but not considered further in the following. Since these two sectors
are interchanged by $T$, the modular invariant partition function
is forced to take the form
  $$Z = a(Z^{{\rm NS}} + Z^{\widetilde{{\rm NS}}} + Z^{{\rm R}})
      + bZ^{\widetilde{{\rm R}}}\,,\eqno(6.4)$$
where $Z^{{\rm A}}$ denotes the diagonal partition function of the characters
of the A-sector, i.e.\
$Z^{{\rm A}} = \sum_{\lambda}\left\mymid\chi^{\sw}_{\lambda,{\rm A}}(\tau)
 \right\mymid^2$.
Here $a,b$ are free constants up to normalization, and
$Z^{\widetilde{{\rm R}}}$ is nothing else than ${\rm tr}(-)^F$. This particular
ansatz cancels out the fermionic contributions in the NS-sector leaving
us with the bosonic characters
${\rm tr}_{\vac{h}}(1 + (-)^F)q^{L_{0}-c/24}$. Thus, partition function
and characters are divided in the same sectors (of (anti-) periodic
boundary conditions) as in the case of the ADE-classification
of the minimal theories of the supersymmetric Virasoro-algebra by
Cappelli \q{7}\q{8}.
Note also, that in the R-sector one has
a non-trivial algebra of the zero modes of the fields, which can
involve $2^n$-dimensional additional representations of the Clifford
algebra as for example the representation $(-)^F = \pm 1$ for the
fermion number $F$.\par
  \sn Without loss of generality let us assume $k \in \BN$ (this case we call
the bosonic one in analogy to the $\wa{3k}$-algebras). In this case the
characters are generically -- up
to the appropriate linear combinations of the theta-functions to
different moduli, if the $q$-powers differ by integers -- given by the
functions
  $$\begin{array}{lcll}
  \Lambda_{\lambda,k+\frac{\varepsilon}{2}}^{{\rm NS}}(\tau) &=& {\ds
  \frac{\eta(\frac{\tau+1}{2})}{\eta^2(\tau)}e^{-\frac{\pi i}{24}}
  \Theta_{\lambda,k+\frac{\varepsilon}{2}}(\tau)}
  & {\rm NS-sector}\,,\strut{-3ex}{3ex}\\
  \Lambda_{\lambda+\frac{\varepsilon}{2},k+\frac{\varepsilon}{2}}^{{\rm
  R}}(\tau) &=& {\ds
  \frac{\eta(2\tau)}{\eta^2(\tau)}
  \Theta_{\lambda+\frac{1}{2},k+\frac{\varepsilon}{2}}(\tau)}
  & {\rm R-sector}\,,\strut{0ex}{4ex}\end{array}\eqno(6.5)$$
where $\lambda\in\BZ$ and again $\varepsilon = 0$ or $1$. Note that
$TST$ intertwines both sectors. All weights in the R-sector have to be
shifted by $\frac{1}{16}$. Last but not least one again has to deal
with representations with multiplicities greater one, too, if the
eigenvalue of the second element of the Cartan subalgebra does not
vanish. The following table sums up our results.
  $$\begin{array}{ll|c||rcl|ll}
  \multicolumn{2}{l}{{\rm Table\ 6.2\ }}
    & \multicolumn{6}{l}{{\scriptstyle{\rm
    representations\ and\ their\ characters\ for\ the\ bosonic\ }
    \swa{3k}{\rm -algebras}}} \\ \hline
  \phantom{xxx} & \multicolumn{6}{l}{\phantom{-}} & \\
  & h \strut{-1ex}{1ex}                                & w^2
    & \chi_{\lambda}^{\sw}                             &
    &
    & {\scriptstyle{\rm remark}}                       & \\ \cline{2-7}
NS:&h_{1,1} \strut{-1ex}{4ex}                          & 0
    & \chi_{0,{\rm NS}}^{\sw}                          & =
    & \frac{1}{2}(\Lambda_{0,k}^{{\rm NS}} - \Lambda_{0,k+1/2}^{{\rm NS}})
    & {\scriptstyle{\rm vacuum\ rep.}}                 & \\
  & h_{\frac{1}{2k},\frac{1}{2k}} \strut{-1ex}{1ex}    & 0
    & \chi_{1,{\rm NS}}^{\sw}                          & =
    & \Lambda_{1,k}^{{\rm NS}}
    &                                                  & \\
  & \vdots                                             & \vdots
    &                                                  & \vdots
    &
    &                                                  & \\
  & h_{\frac{k-1}{2k},\frac{k-1}{2k}} \strut{-1ex}{1ex}& 0
    & \chi_{k-1,{\rm NS}}^{\sw}                        & =
    & \Lambda_{k-1,k}^{{\rm NS}}
    &                                                  & \\
  & h_{\frac{1}{2},\frac{1}{2}} \strut{-1ex}{1ex}      & \neq 0
    & \chi_{k,+,{\rm NS}}^{\sw} = \chi_{k,-,{\rm NS}}^{\sw} & =
    & \frac{1}{2}\Lambda_{k,k}^{{\rm NS}}
    & {\scriptstyle{\rm degenerate\ rep.}}             & \\
  & h_{0,0} \strut{-1ex}{1ex}                          & 0
    & \chi_{k+1,{\rm NS}}^{\sw}                        & =
    & \frac{1}{2}(\Lambda_{0,k}^{{\rm NS}} + \Lambda_{0,k+1/2}^{{\rm NS}})
    & {\scriptstyle{\rm rep.\ to\ }h_{{\em min}}}      & \\
  & h_{\frac{1}{2k+1},-\frac{1}{2k+1}}\strut{-1ex}{1ex}& 0
    & \chi_{-1,{\rm NS}}^{\sw}                         & =
    & \Lambda_{1,k+1/2}^{{\rm NS}}
    &                                                  & \\
  & \vdots \strut{-1ex}{1ex}                           & \vdots
    &                                                  & \vdots
    &
    &                                                  & \\
  & h_{\frac{k}{2k+1},-\frac{k}{2k+1}}\strut{-1ex}{1ex}& 0
    & \chi_{-k,{\rm NS}}^{\sw}                         & =
    & \Lambda_{k,k+1/2}^{{\rm NS}}
    &                                                  & \\ \cline{2-7}
 R:&h_{0,0}+\frac{1}{16} \strut{-1ex}{4ex}             & 0
    & \chi_{0,{\rm R}}^{\sw}                           & =
    & \frac{1}{2}\Lambda_{0,k}^{{\rm R}}
    &                                                  & \\
  & h_{\frac{1}{2k},\frac{1}{2k}}+\frac{1}{16} \strut{-1ex}{1ex} & 0
    & \chi_{1,{\rm R}}^{\sw}                           & =
    & \Lambda_{1,k}^{{\rm R}}
    &                                                  & \\
  & \vdots                                             & \vdots
    &                                                  & \vdots
    &
    &                                                  & \\
  & h_{\frac{k-1}{2k},\frac{k-1}{2k}} + \frac{1}{16} \strut{-1ex}{1ex}& 0
    & \chi_{k-1,{\rm R}}^{\sw}                         & =
    & \Lambda_{k-1,k}^{{\rm R}}
    &                                                  & \\
  & h_{\frac{1}{2},\frac{1}{2}} + \frac{1}{16} \strut{-1ex}{1ex} & \neq 0
    & \chi_{k,+,{\rm R}}^{\sw} = \chi_{k,-,{\rm R}}^{\sw} & =
    & \frac{1}{2}\Lambda_{k,k}^{{\rm R}}
    & {\scriptstyle{\rm degenerate\ rep.}}             & \\
  & h_{\frac{1}{4k+2},-\frac{1}{4k+2}} + \frac{1}{16} \strut{-1ex}{1ex} & 0
    & \chi_{-1,{\rm R}}^{\sw}                          & =
    & \Lambda_{1/2,k+1/2}^{{\rm R}}
    &                                                  & \\
  & h_{\frac{3}{4k+2},-\frac{3}{4k+2}} + \frac{1}{16} \strut{-1ex}{1ex} & 0
    & \chi_{-2,{\rm R}}^{\sw}                          & =
    & \Lambda_{3/2,k+1/2}^{{\rm R}}
    &                                                  & \\
  & \vdots \strut{-1ex}{1ex}                           & \vdots
    &                                                  & \vdots
    &
    &                                                  & \\
  & h_{\frac{2k-1}{2k+1},-\frac{2k-1}{2k+1}} + \frac{1}{16} \strut{-1ex}{1ex}&
0
    & \chi_{-k,{\rm R}}^{\sw}                          & =
    & \Lambda_{k-1/2,k+1/2}^{{\rm R}}
    &                                                  & \\
  & h_{\frac{1}{2},-\frac{1}{2}} + \frac{1}{16} \strut{-1ex}{1ex} & \neq 0
    & \chi_{-k-1,+,{\rm R}}^{\sw} = \chi_{-k-1,-,{\rm R}}^{\sw} & =
    & \frac{1}{2}\Lambda_{k+1/2,k+1/2}^{{\rm R}}
    & {\scriptstyle{\rm degenerate\ rep.}}             & \\ \cline{2-7}
  \multicolumn{8}{l}{\phantom{-}} \\ \hline\end{array}$$
In the case of the fermionic $\sw$-algebras, i.e.\ $k\in\BZ_{+}+\frac{1}{2}$,
the r\^{o}le of $k$ and $k+\frac{1}{2}$ interchanges since $k$ is now
half-integer.\par
  \bn As in the ordinary case there exist the so called odd sector algebras
$\swa{8k}$. The characters of the NS-sector are built up from the
functions
  $$\Lambda_{\lambda,4k+2\varepsilon}^{{\rm NS}}(\tau)
  = \frac{\eta\left(\frac{\tau+1}{1}\right)}{\eta^2(\tau)}
  e^{-\frac{\pi i}{24}}\Theta_{\lambda,4k+2\varepsilon}(\tau)\,,\eqno(6.6)$$
where several linear combinations occur analogous to the characters of
the $\wa{8k}$-algebras. In the R-sector we have to use the functions
  $$\Lambda_{\lambda,4k+2\varepsilon}^{{\rm R}}(\tau)
  = \frac{\eta(2\tau)}{\eta^2(\tau)}
  \Theta_{\lambda,4k+2\varepsilon}(\tau)\,,\eqno(6.7)$$
where no linear combinations are possible due to the different parity
of $h_{r,r}$ and $h_{r,-r}$ with respect to the $(-)^F$-operator representation
appearing in the Ramond sector. But the lowest-weight representations
to $h_{1,1}$, $h_{0,0}$ and $h_{1,-1}$ are now each twofold degenerate,
there are two values for the eigenvalue $w$ of $\Phi_0 = \Phi^{(3)}_0$.
In our case we have two representations
at the ground level $h_{{\em min}}$. Some higher level representations
can now built up on either of these ground state representations by
applying the mode $\Phi^{(2)}_{\frac{1}{2}}$ for $k \in \BZ_{+} + \frac{1}{2}$
or the mode $\Phi^{(2)}_1$ for $k \in \BZ_{+}$ on these ground states.
(Note that for $k \in \BZ_{+} + \frac{1}{4}$ this cannot be understood in the
frame of chiral $\sw$-algebras, since this field is not local to itself, but
to the other local fields of the $\sw$-algebra, $\BI$, $\Phi^{(3)}$, etc.
Therefore the action of the field $\Phi^{(2)}$ does not cause a real
problem, as long as only one mode (symbolically notation!)
$\Phi^{(2)}_{\frac{1}{4}}$ is allowed to appear in the monomials
of the mode expansion of the resulting state.)
The following table lists all irreducible lowest-weight representations:
  $$\begin{array}{ll|c||rcl|ll}
  \multicolumn{2}{l}{{\rm Table\ 6.3\ }}
    & \multicolumn{6}{l}{{\scriptstyle{\rm
    representations\ and\ their\ characters\ for\ the\ odd\ sector\ }
    \swa{8k}{\rm -algebras}}} \\ \hline
  \phantom{xxx} & \multicolumn{6}{l}{\phantom{-}} & \\
  & h \strut{-1ex}{1ex}                                & w
    & \chi_{\lambda,{\em odd}}^{\sw}                   &
    &
    & {\scriptstyle{\rm remark}}                       & \\ \cline{2-7}
NS:&h_{1,1} \strut{-1ex}{4ex}                          & 0
    & \chi_{0,{\rm NS}}^{\sw}                          & =
    & \frac{1}{2}(\Lambda_{4k,4k}^{{\rm NS}} - \Lambda_{4k,4k+2}^{{\rm NS}})
    & {\scriptstyle{\rm vacuum\ rep.}}                 & \\
  & h_{\frac{1}{4k},\frac{1}{4k}} \strut{-1ex}{1ex}    & \neq 0
    & \chi_{1,{\rm NS}}^{\sw}                          & =
    & \Lambda_{1+4k,4k}^{{\rm NS}}
    &                                                  & \\
  & \vdots                                             & \vdots
    &                                                  & \vdots
    &
    &                                                  & \\
  & h_{\frac{4k-1}{4k},\frac{4k-1}{4k}} \strut{-1ex}{1ex}& \neq 0
    & \chi_{4k-1,{\rm NS}}^{\sw}                       & =
    & \Lambda_{8k-1,4k}^{{\rm NS}} = \Lambda_{-1,4k}^{{\rm NS}}
    &                                                  & \\
  & h_{2,2} \strut{-1ex}{1ex}                          & \neq 0
    & \chi_{4k,{\rm NS}}^{\sw}                         & =
    & \frac{1}{2}(\Lambda_{0,4k}^{{\rm NS}} - \Lambda_{0,4k+2}^{{\rm NS}})
    & {\scriptstyle{\rm rep.\ on\ }\vac{\Phi^{(2)}}}   & \\
  & h_{0,0} \strut{-1ex}{1ex}                          & \neq 0
    & \chi_{4k+1,{\rm NS}}^{\sw}                       & =
    & \frac{1}{2}(\Lambda_{0,4k}^{{\rm NS}} + \Lambda_{0,4k+2}^{{\rm NS}})
    & {\scriptstyle{\rm rep.\ to\ }h_{{\em min}}}      & \\
  & h_{1,1} \strut{-1ex}{1ex}                          & \neq 0
    & \chi_{4k+2,{\rm NS}}^{\sw}                       & =
    & \frac{1}{2}(\Lambda_{4k,4k}^{{\rm NS}} + \Lambda_{4k,4k+2}^{{\rm NS}})
    & {\scriptstyle\vac{h=0,w\neq 0}{\rm\ rep.}}       & \\
  & h_{\frac{1}{4k+2},-\frac{1}{4k+2}}\strut{-1ex}{1ex}& \neq 0
    & \chi_{-1,{\rm NS}}^{\sw}                         & =
    & \Lambda_{1+4k+2,4k+2}^{{\rm NS}}
    &                                                  & \\
  & \vdots \strut{-1ex}{1ex}                           & \vdots
    &                                                  & \vdots
    &
    &                                                  & \\
  & h_{\frac{4k+1}{4k+2},-\frac{4k+1}{4k+2}}\strut{-2ex}{2ex}& \neq 0
    & \chi_{-4k-1,{\rm NS}}^{\sw}                      & =
    & \Lambda_{8k+3,4k+2}^{{\rm NS}} = \Lambda_{-1,4k+2}^{{\rm NS}}
    &                                                  & \\ \cline{2-7}
 R:&h_{1,1}+\frac{1}{16} \strut{-1ex}{4ex}             & w_1\neq w_2
    & \chi_{0,+,{\rm R}}^{\sw}                         & =
    & \frac{1}{2}\Lambda_{4k,4k}^{{\rm R}}
    & {\scriptstyle{\rm degenerate\ rep.}}             & \\
  & h_{\frac{1}{4k},\frac{1}{4k}}+\frac{1}{16}\strut{-1ex}{1ex} & \neq 0
    & \chi_{1,{\rm R}}^{\sw}                           & =
    & \Lambda_{1+4k,4k}^{{\rm R}}
    &                                                  & \\
  & \vdots                                             & \vdots
    &                                                  & \vdots
    &
    &                                                  & \\
  & h_{\frac{4k-1}{4k},\frac{4k-1}{4k}}+\frac{1}{16}\strut{-1ex}{1ex}& \neq 0
    & \chi_{4k-1,{\rm R}}^{\sw}                        & =
    & \Lambda_{8k-1,4k}^{{\rm R}} = \Lambda_{-1,4k}^{{\rm R}}
    &                                                  & \\
  & h_{0,0}+\frac{1}{16} \strut{-1ex}{1ex}             & \neq 0
    & \chi_{4k,{\rm R}}^{\sw}                          & =
    & \frac{1}{2}\Lambda_{0,4k}^{{\rm R}}
    & {\scriptstyle{\rm first\ rep.\ to\ }h_{{\em min}}}& \\
  & h_{1,-1}+\frac{1}{16} \strut{-1ex}{1ex}            & w_1\neq w_2
    & \chi_{0,-,{\rm R}}^{\sw}                         & =
    & \frac{1}{2}\Lambda_{4k+2,4k+2}^{{\rm R}}
    & {\scriptstyle{\rm degenerate\ rep.}}             & \\
  & h_{\frac{1}{4k+2},-\frac{1}{4k+2}}+\frac{1}{16}\strut{-1ex}{1ex}& \neq 0
    & \chi_{-1,{\rm R}}^{\sw}                          & =
    & \Lambda_{1+4k+2,4k+2}^{{\rm R}}
    &                                                  & \\
  & \vdots \strut{-1ex}{1ex}                           & \vdots
    &                                                  & \vdots
    &
    &                                                  & \\
  & h_{\frac{4k+1}{4k+2},-\frac{4k+1}{4k+2}}+\frac{1}{16}
    \strut{-2ex}{2ex} & \neq 0
    & \chi_{-4k-1,{\rm R}}^{\sw}                       & =
    & \Lambda_{8k+3,4k+2}^{{\rm R}} = \Lambda_{-1,4k+2}^{{\rm R}}
    &                                                  & \\
  & h_{0,0}+\frac{1}{16} \strut{-1ex}{1ex}             & \neq 0
    & \chi_{-4k-2,{\rm R}}^{\sw}                       & =
    & \frac{1}{2}\Lambda_{0,4k+2}^{{\rm R}}
    & {\scriptstyle{\rm second\ rep.\ to\ }h_{{\em min}}}& \\ \cline{2-7}
  \multicolumn{8}{l}{\phantom{-}} \\ \hline\end{array}$$
  \par\bn In complete analogy to chapter 3 the structure constants and
decomposition coefficients into chiral BRST-invariant vertex operators
can be calculated. In the papers \q{1}\q{28} the supersymmetric
versions of the normalization integerals of Dotsenko-Fateev type are
calculated. There it is shown, that the monodromy coefficients are
given by the same formulae as in the conformal case, only $\alpha_{+}^2$
has to be substituted by $\frac{\alpha_{+}^{2}-1}{2}$. But this means
that the same is true for the braid matrices of \q{17} and
consequently for the $\Delta$-coefficients in (3.2). Using this and
the normalization constants
  $$N_{(k'k)(l'l)}^{(p'p)} = \widehat{I}_{r',r}(\alpha_{+}\alpha_{k',k},
  \alpha_{+}\alpha_{l',l},\alpha_{+}^2)\eqno(6.8)$$
taken from \q{1}, where $p = k+l-2r-1$ and similar for $p'$,
we finally obtain in our case ($r = r'$, thus in particular $r+r'$ even)
for the self-coupling structure constants (actually their square)
  $$\left(C_{\Phi\Phi}^{\Phi}\right)^2 = \left\{
  \begin{array}{l}
    {\ds\frac{(\frac{3}{2}-24k)}{8k}
        \frac{\prod_{j=1}^{8k}\left(j^2-64(k^2+\frac{k}{2})\right)^2
              \prod_{j=1}^{2k}\left(j^2-4(k^2+\frac{k}{2})\right)^3}
             {\prod_{j=1}^{6k}\left(j^2-36(k^2+\frac{k}{2})\right)
              \prod_{j=1}^{4k}\left(j^2-16(k^2+\frac{k}{2})\right)^4}}
    \strut{-5ex}{5ex}\\
    \phantom{--------}{\rm if}\ k\in\BZ_{+}/2\,,
    \strut{-2ex}{3ex}\\
    {\ds\frac{(\frac{3}{2}-24k)}{8k}
        \frac{\prod_{j=1}^{8k}\left(j^2-64(k^2+\frac{k}{2})\right)^2
              \prod_{j=1}^{2k+1/2}\left((j-\frac{1}{2})^2
                -4(k^2+\frac{k}{2})\right)^3}
             {\prod_{j=1}^{6k+1/2}\left((j-\frac{1}{2})^2
                -36(k^2+\frac{k}{2})\right)
              \prod_{j=1}^{4k}\left(j^2-16(k^2+\frac{k}{2})\right)^4}
        \frac{3}{4(k^2+\frac{k}{2})}}
    \strut{-5ex}{5ex}\\
    \phantom{--------}{\rm if}\ k\in\BZ_{+}+\frac{1}{4}\,.
  \end{array}\right.\eqno(6.9)$$
Needless to say that these results explain all explicit calculations
of these algebras obtained so far. The supersymmetric case shows a
structure, which is closely related to the conformal case, the only
surprise coming from the Ramond sector. Hence we do not want to be
more detailed here.
  \bn\bn\bn
  \smallkapitel{7}{Summary and Conclusion}{}
  \smallcapitel{W}ith this paper we established a whole class of new
RCFTs which are not related to minimal models or any coset constructions.\par
  \sn First, starting from some explicitly calculated examples \q{5} we
constructed a class of extended chiral symmetry algebras related to the
non-minimal Dotsenko-Fateev models \q{12}. In these models for the
special values $c = 1 - 24k$, $k\in\BN/4$ for the central charge the
requirement of
locality for the chiral symmetry algebra enables one to determine the
field content of the latter, which turns out to be a finitely generated
$\w$-algebra, and to give abstract fusion rules.\par
  \sn Secondly, we were able to calculate the non-trivial structure
constants of these $\w$-algebras, namely the self-coupling of the additional
primary field, and, as a byproduct, the decomposotion coefficients of this
field into its chiral BRST-invariant vertex operators. The results are
strongly related to the expressions obtained by Felder and Fr\"ohlich
\q{17} from the braiding properties of the chiral vertex
operators. By a redefinition of the chiral vertex operators and of the
screening charges the algebra gets a thermal structure simplifying the
braid group representation and the analytical behaviour of the
Dotsenko-Fateev integrals.\par
  \sn Thirdly, we calculated the characters of the vacuum representations
of these chiral algebras and then, via modular transformations, the
indeed finite set of all representations yielding the complete CFT which
turns out to be rational. Thus, the chiral algebra already is maximally
extended since it diagonalizes the modular invariant partition function.
In particular, we worked out
the explicit form of the $S$-matrix and the structure constants of the
fusion algebra for the subclass of bosonic $\w$-algebras. The fermionic
case and the case of the odd sector subalgebras were briefly discussed.
All results are in complete agreement with the explicitly calculated
examples in \q{13}.\par
  \sn Next, we completed the proof of the classification of all
RCFTs with central charge $c = 1$, given by Kiritsis \q{27},
towards the non-unitary case. For this we used
the effective central charge $c_{{\em eff}} = c - 24h_{{\em min}}$.
It turned out that the models discribed in this work are the only
possible non-unitary ones, who have $c_{{\em eff}} = 1$. They fit in
the only case of a modular invariant partition function, which
had been rejected by Kiritsis due to his restriction to unitary
theories.\par
  \sn In the last part of the paper we outlined a generalization of
our results to the supersymmetric case where also some examples of
$\cal{S}\w$-algebras are now available, see \q{4} and \q{14}
for their representations. The structure of the results is very
similar to the non supersymmetric case.\par
  \sn Our arguments cover the complete set of possible chiral extended
symmetry algebras and thus RCFTs coming from degenerate models, since
these are either minimal models and coset constructions or the models
discussed in this work. In particular, the classification of all RCFTs
with $c_{\em eff} = 1$ is completed including the non-unitary case.\par
  \bn Still,
a lot of questions remain open. The most exciting one in the frame of this
work might be, what the other possible combinations of theta-functions
with moduli say $k$ and $k+k'$, $k'\not\in \{0,\frac{1}{2},1,2,4\}$
physically
could mean. In our case the combinations were necessary due to the embedding
structure of the Virasoro Verma-modules coming from null states.
Let us again stress the point of $c$ rational but not contained in the
minimal series nor in the set $c = 1 - 24k$, $k\in\BN/4$. From equation
(4.18) one might think that there should be at least
RCFTs for $k = \frac{p}{\alpha^2}$.
With this ansatz one obtains (again $\varepsilon = 0$ or $1$)
  $$\Lambda_{\lambda,\alpha^{2}(k+\varepsilon)}(\tau)
  = \frac{1}{\eta(\tau)} \sum_{n\in\BZ}
  q^{(\alpha n + \frac{\alpha\lambda}{2\alpha^{2}(k+\varepsilon)})^{2}
  (k+\varepsilon)} = \frac{1}{\eta(\tau)}q^{k}
  \sum_{r\in\alpha\BZ+\frac{\alpha\lambda}{2\alpha^{2}(k+\varepsilon)}}
  q^{h_{r,(-)^{\varepsilon}r}}\,,\eqno(7.1)$$
which yields a condition on $r$ or equivalently on $\lambda$
in order to get integer or half-integer weights
  $$h_{r,(-)^{\varepsilon}r} = \left(\left(\alpha n + \frac{\alpha\lambda}
  {2\alpha^{2}(k+\varepsilon)}\right)^2 - 1\right)\frac{p}{\alpha^2}
  = pn^2 + n\lambda + \frac{\lambda^2}{4p} - \frac{p}{\alpha^2}\,.
  \eqno(7.2)$$
Therefore we must put $\lambda = p = \alpha^{2}(k+\varepsilon)$
resulting in the condition
$\frac{p}{4} - \frac{p}{\alpha^2} \in \BZ$ which can only be fullfilled
for $\alpha = 2$ corresponding to our odd sector subalgebras. Thus,
these algebras are the only ones which can be extracted out of a larger
set of not necessarily chiral local operators. This is the case for
$k\in\BN/4$, where the even operators are not local to themselves and hence
cannot be added to the chiral algebra.\par
  \sn Another
question might be, whether the labeling of the lowest-weight levels $h$
with rational indices has some physical meaning in the frame of
$\w$-gravity, where e.g.\ rational powers of screening charges are used
\q{21}\q{11}.\par
  \sn Finally, the classification of all RCFTs,
in particular for $c > 1$, is still far away from being completed.
But a big step towards the classification of all $\w$-algebras with
one additional generator could be achieved. The situation is now the
following: Several classes of such $\w(2,\delta)$-algebras have been
established.
\begin{list}{}{}
  \item[(i)] The generically existing algebras $\w(2,\delta)$ with
    $\delta \in \{\frac{1}{2},1,\frac{3}{2},2,3,4,6\}$. All these
    algebras have well known classical counterparts as the algebra
    of Casimir operators of a Lie-algebra.
  \item[(ii)] The algebras, which exist for $c$ an element of the
    minimal series $c = 1 - 6\frac{(p-q)^2}{pq}$, $p,q\in\BN$ coprime.
    These algebras are related to the ADE-classification of modular
    invariant partition functions of Virasoro minimal models \q{8},
    as has been worked out in \q{36} for the fermionic case.
  \item[(iii)] $\wa{2q-1}$-algebras to $c = 1 - 6\frac{(1-q)^2}{q}$, $q\in\BN$,
    (minimal series with $p = 1$). These algebras have been studied in \q{25}.
    They are not extended symmetry algebras of a RCFT.
  \item[(iv)] The $\wa{3k}$ and $\wa{8k}$ algebras with central charge
    $c = 1 - 24k$, $k\in\BN/4$ as discussed in this paper. These are the
    only algebras related to non-minimal degenerate Virasoro models.
  \item[(v)] $\w$-algebras to isolated irrational $c$-values. Following
    \q{2}, these algebras cannot belong to RCFTs.
  \item[(vi)] Some exceptional $\w$-algebras, mainly the $\w(2,8)$ for all
    the values of the central charge not covered by (i) to (v), have
    been found. Probably they are related to other finite groups
    which can be represented by the modular group, see \q{13}.
    In all these cases the self-coupling is non zero.
  \end{list}\noindent
A very similar pattern is valid for the $\swa{\delta}$-algebras (but
without solutions of type (v), i.e.\ without isolated irrational solutions).
Since all known examples of set (vi) have non-vanishing self-coupling,
we conjecture that the classification of the $\wa{\delta/2}$-algebras
to rational central charge, $\delta\in\BN$, is complete.
  \par\sn We
want to conclude with one very speculative remark. We have found RCFTs
for the central charges of the form $c = 1 - x$ with $x$ having divisor
$24$ (bosonic case) or $12$ (fermionic case). There is an interesting
work by Goddard \q{10} in which nice RCFTs, related to selfdual
even lattices, with central charges $c = x$, $x$ having divisor $24$ (bosonic
case) or $12$ (fermionic case) are found. Is there a relation between
these theories or even in general between theories with $c$ and $1 - c$?\par
  \vfill
  \bn I would like to thank R.\ Blumenhagen, A.\ Recknagel, M.\ Terhoeven
and in particular W.\ Eholzer, A.\ Honecker R.\ H\"ubel,
and R.\ Varnhagen for a lot of discussions and comments. I am extremely
grateful to W.\ Nahm, in particular for most valuable discussions on chapter
five. Last but not least I thank the Deutsche Forschungsgemeinschaft for
financial support.
  \newpage\noindent
  \smallkapitel{Appendix A}{The S and T matrix for {\boldmath $\w(2,3k)$}}{}
  \smallcapitel{T}his Appendix presents the general form of the S-matrix
for the case of bosonic $\w(2,3k)$ theories, i.e.\ $k\in\BN$. For the
Neveu-Schwarz sector of the fermionic case ($k\in\BN+\frac{1}{2}$) the
$S$-matrix is exactly the same, if one removes the rows and columns
belonging to the characters $\chi_{k,\pm}^{\w}$ and $\chi_{-k-1,\pm}^{\w}$,
i.e.\ to the degenerate representations, and if one substitutes
$k$ by $\frac{k}{2}$. So, let $k\in\BN$. Define the functions
${\cal C}_{\alpha}(x) = \frac{1}{\sqrt{2\alpha}}\cos(\pi\frac{x}{\alpha})$.
Then $S$ is given by
\def\C#1#2{{\cal C}_{#1}(#2)}
{\footnotesize
  $$\begin{array}{l}
  \left(\begin{array}{cccccc}
    \frac{1}{2}\left(\C{k}{0}+\C{k+1}{0}\right) & \C{k}{0}            &
    \cdots                                      & \C{k}{0}            &
    \frac{1}{2}\C{k}{0}                         & \frac{1}{2}\C{k}{0}
  \strut{-1ex}{1ex}\\
    \C{k}{0}                                    & 2\C{k}{1}           &
    \cdots                                      & 2\C{k}{k-1}         &
    \C{k}{k}                                    & \C{k}{k}
  \strut{-1ex}{1ex}\\
    \vdots                                      & \vdots              &
                                                & \vdots              &
    \vdots                                      & \vdots
  \strut{-1ex}{1ex}\\
    \C{k}{0}                                    & 2\C{k}{k-1}         &
    \cdots                                      & 2\C{k}{(k-1)^2}     &
    \C{k}{k(k-1)}                               & \C{k}{k(k-1)}
  \strut{-1ex}{1ex}\\
    \frac{1}{2}\C{k}{0}                         & \C{k}{k}            &
    \cdots                                      & \C{k}{k(k-1)}       &
    A                                           & \C{k}{k^2} - A
  \strut{-1ex}{1ex}\\
    \frac{1}{2}\C{k}{0}                         & \C{k}{k}            &
    \cdots                                      & \C{k}{k(k-1)}       &
    \C{k}{k^2} - A                              & A
  \strut{-1ex}{1ex}\\
    \frac{1}{2}\left(\C{k}{0}-\C{k+1}{0}\right) & \C{k}{0}            &
    \cdots                                      & \C{k}{0}            &
    \frac{1}{2}\C{k}{0}                         & \frac{1}{2}\C{k}{0}
  \strut{-1ex}{1ex}\\
    -\C{k+1}{0}                                 & 0                   &
    \cdots                                      & 0                   &
    0                                           & 0
  \strut{-1ex}{1ex}\\
    \vdots                                      & \vdots              &
                                                & \vdots              &
    \vdots                                      & \vdots
  \strut{-1ex}{1ex}\\
    -\C{k+1}{0}                                 & 0                   &
    \cdots                                      & 0                   &
    0                                           & 0
  \strut{-1ex}{1ex}\\
    -\frac{1}{2}\C{k+1}{0}                      & 0                   &
    \cdots                                      & 0                   &
    C                                           & -C
  \strut{-1ex}{1ex}\\
    -\frac{1}{2}\C{k+1}{0}                      & 0                   &
    \cdots                                      & 0                   &
     -C                                         & C
  \end{array}\right.\\
  \phantom{-}\\
  \left.\begin{array}{cccccc}
    \frac{1}{2}\left(\C{k}{0}-\C{k+1}{0}\right) & -\C{k+1}{0}         &
    \cdots                                      & -\C{k+1}{0}         &
    -\frac{1}{2}\C{k+1}{0}                      & -\frac{1}{2}\C{k+1}{0}
  \strut{-1ex}{1ex}\\
    \C{k}{0}                                    & 0                   &
    \cdots                                      & 0                   &
    0                                           & 0
  \strut{-1ex}{1ex}\\
    \vdots                                      & \vdots              &
                                                & \vdots              &
    \vdots                                      & \vdots
  \strut{-1ex}{1ex}\\
    \C{k}{0}                                    & 0                   &
    \cdots                                      & 0                   &
    0                                           & 0
  \strut{-1ex}{1ex}\\
    \frac{1}{2}\C{k}{0}                         & 0                   &
    \cdots                                      & 0                   &
    C                                           & -C
  \strut{-1ex}{1ex}\\
    \frac{1}{2}\C{k}{0}                         & 0                   &
    \cdots                                      & 0                   &
    -C                                          & C
  \strut{-1ex}{1ex}\\
    \frac{1}{2}\left(\C{k}{0}+\C{k+1}{0}\right) & \C{k+1}{0}          &
    \cdots                                      & \C{k+1}{0}          &
    \frac{1}{2}\C{k+1}{0}                       & \frac{1}{2}\C{k+1}{0}
  \strut{-1ex}{1ex}\\
    \C{k+1}{0}                                  & 2\C{k+1}{1}         &
    \cdots                                      & 2\C{k+1}{k}         &
    \C{k+1}{k+1}                                & \C{k+1}{k+1}
  \strut{-1ex}{1ex}\\
    \vdots                                      & \vdots              &
                                                & \vdots              &
    \vdots                                      & \vdots
  \strut{-1ex}{1ex}\\
    \C{k+1}{0}                                  & 2\C{k+1}{k}         &
    \cdots                                      & 2\C{k+1}{k^2}       &
    \C{k+1}{k(k+1)}                             & \C{k+1}{k(k+1)}
  \strut{-1ex}{1ex}\\
    \frac{1}{2}\C{k+1}{0}                       & \C{k+1}{k+1}        &
    \cdots                                      & \C{k+1}{k(k+1)}     &
    B                                           & \C{k+1}{(k+1)^2} - B
  \strut{-1ex}{1ex}\\
    \frac{1}{2}\C{k+1}{0}                       & \C{k+1}{k+1}        &
    \cdots                                      & \C{k+1}{k(k+1)}     &
    \C{k+1}{(k+1)^2} - B                        & B
  \end{array}\right)\,.\end{array}$$
}{\raggedleft(A.1)\par}\sn                   
Here the three free parameters are determined by the requirement that
the fusion algebra structure constants are non-negative integers. This
is carried out in appendix B. The result is
  $$\begin{array}{lcl}
    C & = & {\ds ( i)^{k  }      \frac{1}{\sqrt{8}}                   }
      \,,\strut{-3ex}{3ex}\\
    A & = & {\ds (-1)^{k  }\left(\frac{1}{2\sqrt{2k}}   \pm C \right) }
      \,,\strut{-3ex}{7ex}\\
    B & = & {\ds (-1)^{k+1}\left(\frac{1}{2\sqrt{2k+2}} \pm C \right) }
      \,.\strut{0ex}{4ex}
  \end{array}\eqno(A.2)$$
Note that $\frac{1}{\sqrt{2k}}(-1)^{k} - A_{\pm}
= (-1)^{k}(\frac{1}{2\sqrt{2k}} \mp C) = A_{\mp}$ and similarly for $B$.
Thus the two solutions for $A$ and $B$ just mean an interchange or a
reordering of the degenerate representations.
The $T$-matrix is much simpler and given by
  $$\begin{array}{rl}{\rm diag}& \!\!\!\!\!\left(
    \exp(\pi i(                 - \frac{1}{12})),
    \exp(\pi i(\frac{1}{2k}     - \frac{1}{12})),\ldots,
    \exp(\pi i(\frac{k}{2k}     - \frac{1}{12})),
    \exp(\pi i(\frac{k}{2k}     - \frac{1}{12})),
  \right.\strut{-3ex}{3ex} \\  & \left.
    \exp(\pi i(                 - \frac{1}{12})),
    \exp(\pi i(\frac{1}{2k+2}   - \frac{1}{12})),\ldots,
    \exp(\pi i(\frac{k+1}{2k+2} - \frac{1}{12})),
    \exp(\pi i(\frac{k+1}{2k+2} - \frac{1}{12}))
  \right)\,.\end{array}\eqno(A.3)$$
The $S$ and $T$ matrices for the odd sector algebras are easy to obtain from
equations (4.5) and table 4.2, which shows how the characters can be
expressed in $\Lambda$-functions. Therefore we do not go into further
details here.
  \bn\bn\bn
  \smallkapitel{Appendix B}{The Fusion Algebra for {\boldmath $\w(2,3k)$}}{}
  \smallcapitel{F}inally, in this appendix we calculate the fusion algebra
for the bosonic $\w(2,3k)$ and show explicitly that all structure
constants are indeed non-negative integers. This also determines the
free paramters of the extended $S$-matrix uniquely completing
the proof of rationality of the theories.\par
  \sn One starts with $A,B,C\in\BC$ as arbitrary free complex
numbers. With the Verlinde formula one calculates some particular
structure constants $N_{\alpha\beta}^{\gamma}$. For example
  $$\begin{array}{lclclcl}
    {\ds N_{k,+;k,+}^{-j}} & = & {\ds\frac{1}{2} - (-1)^{j}4C^{2}}\,, &
      \phantom{---}\strut{-2ex}{2ex} &
    {\ds N_{k,+;k,-}^{-j}} & = & {\ds\frac{1}{2} + (-1)^{j}4C^{2}}\,, \\
    {\ds N_{k,+;-j}^{k,+}} & = & {\ds\frac{1}{2} - (-1)^{j}4\mymid C\mymid^{2}}
      \,,& \phantom{---} \strut{0ex}{3ex} &
    {\ds N_{k,+;-j}^{k,-}} & = & {\ds\frac{1}{2} + (-1)^{j}4\mymid C\mymid^{2}}
      \,,
  \end{array}\eqno(B.1)$$
where $-k\leq -j\leq-1$. Since all these numbers should be non-negative
integral ones, the only solutions are either $0$ or $1$. Hence we get
$\mymid C\mymid^2 = \pm C^2$, i.e.\ $C$ purely real or purely imaginary.
Furthermore, the absolute value is fixed to be $\mymid C\mymid =
\frac{1}{\sqrt{8}}$. This leaves us with the ansatz
  $$C = (i)^{\alpha_C}\frac{1}{\sqrt{8}}\,.\eqno(B.2a)$$
Next we look at the structure constants
  $$N_{k,+;k,-}^{-k-1,+} = N_{k,+;k,-}^{-k-1,-} = \frac{1}{4} + (-1)^{k+1}
  2C^{2}\,.\eqno(B.3)$$
With equation (B.2a) we see that $0$ is the only allowed solution and
therfore we need $\alpha_C \equiv k$ mod $2$. In the following we put
without loss of generality
  $$C = (i)^{k}\frac{1}{\sqrt{8}}\,.\eqno(B.2b)$$\par
  \sn Thirdly, we can consider the following structure constants:
  $$\begin{array}{lcl}
  {\ds N_{k   ,+;k   ,-}^{j  }} & = & {\ds\frac{1}{2} + (-1)^{j}\left(
    4A^2 - \frac{4}{\sqrt{2k  }}(-1)^{k  }A + \frac{1}{2k  }\right)}
    \,,\strut{-3ex}{3ex}\\
  {\ds N_{k   ,+;k   ,+}^{j  }} & = & {\ds\frac{1}{2} - (-1)^{j}\left(
    4A^2 - \frac{4}{\sqrt{2k  }}(-1)^{k  }A + \frac{1}{2k  }\right)}
    \,,\strut{-3ex}{7ex}\\
  {\ds N_{-k-1,+;-k-1,+}^{-j'}} & = & {\ds\frac{1}{2} - (-1)^{j'}\left(
    4B^2 - \frac{4}{\sqrt{2k+2}}(-1)^{k+1}B + \frac{1}{2k+2}\right)}
    \,,\strut{-3ex}{7ex}\\
  {\ds N_{-k-1,+;-k-1,-}^{-j'}} & = & {\ds\frac{1}{2} + (-1)^{j'}\left(
    4B^2 - \frac{4}{\sqrt{2k+2}}(-1)^{k+1}B + \frac{1}{2k+2}\right)}
    \,,\strut{0ex}{4ex}
  \end{array}\eqno(B.4)$$
where $1\leq j\leq k-1$ and $-k\leq -j'\leq 1$. These pairs of
equations again have as only allowed solutions that one equation of a
pair is $1$ and the other $0$. With this information one can solve
the quadratic equations and gets as ansatz
  $$\begin{array}{lcl}
  A & = & {\ds(-1)^{k  }\left(\frac{1}{2\sqrt{2k  }}
    + (i)^{\alpha_A}\frac{1}{\sqrt{8}}\right)}\,,\strut{-3ex}{3ex}\\
  B & = & {\ds(-1)^{k+1}\left(\frac{1}{2\sqrt{2k+2}}
    + (i)^{\alpha_B}\frac{1}{\sqrt{8}}\right)}\,.\strut{0ex}{4ex}
  \end{array}\eqno(B.5)$$
In order to determine the free powers $\alpha_A$ and
$\alpha_B$ one looks at the constants
  $$\begin{array}{lcl}
  {\ds N_{k,-;-k-1,+}^{k,+}}    & = & {\ds \frac{1}{4} - (-1)^{k}
    2\left((A - A^{\ast})C + \mymid C\mymid^2\right)}
    \,,\strut{-2ex}{2ex}\\
  {\ds N_{k,-;-k-1,-}^{k,+}}    & = & {\ds \frac{1}{4} + (-1)^{k}
    2\left((A - A^{\ast})C - \mymid C\mymid^2\right)}
    \,,\strut{-2ex}{5ex}\\
  {\ds N_{k,+;-k-1,-}^{-k-1,+}} & = & {\ds \frac{1}{4} + (-1)^{k+1}
    2\left((B - B^{\ast})C + \mymid C\mymid^2\right)}
    \,,\strut{-2ex}{5ex}\\
  {\ds N_{k,-;-k-1,-}^{-k-1,+}} & = & {\ds \frac{1}{4} - (-1)^{k+1}
    2\left((B - B^{\ast})C - \mymid C\mymid^2\right)}
    \,.\strut{0ex}{3ex}
  \end{array}\eqno(B.6)$$
With equations (B.2) and (B.5) one now sees, that if $C$ is real then
${\rm Im}\, A = {\rm Im}\, B = 0$, or if $C$ is imaginary then
${\rm Im}\, A = -{\rm Im}\, B = \frac{1}{\sqrt{8}}$. Therfore we can
choose $\alpha_A = \alpha_B = k$ obtaining equation (A.2).\par
  \bn Finally we want to list the whole set of the fusion coefficients
to show that indeed they all are non-negative integers. To save space
we only list the not obviously related constants. The others can be
obtained by one of the following formulae: Let $\varepsilon = k\ {\rm mod}\ 2$.
Then we have $S^{2(1+\varepsilon)} = C^{1+\varepsilon} = \BI$. Thus,
the charge conjugation, denoted by $C:\phi_{\alpha} \longmapsto
\phi_{\bar{\alpha}}$ is trivial for $k$ even. Nonetheless we denote
by $E:\phi_{\alpha} \longmapsto \phi_{\widehat{\alpha}}$ the exchange of
the degenerate representations, i.e.\ $\widehat{(k,\pm)} = (k,\mp)$,
analogously for $(-k-1,\pm)$ and $\widehat{\alpha} = \alpha$ else.
Then we have, using the conjugation matrix to raise or lower indices,
  $$\begin{array}{c}
  N_{\alpha\beta}^{\gamma} = N_{\bar{\alpha}\bar{\beta}}^{\bar{\gamma}}
                           = N_{\widehat{\alpha}\widehat{\beta}}^{
                             \widehat{\gamma}}\,,\strut{-2ex}{2ex}\\
  N_{\alpha\beta}^{\gamma} = N_{\alpha\bar{\gamma}}^{\bar{\beta}}
                           = N_{\bar{\gamma}\beta}^{\bar{\alpha}}
                           \,,\strut{-2ex}{5ex}\\
  N_{\alpha\beta}^{\gamma} = N_{\beta\alpha}^{\gamma}
                           \,.\strut{0ex}{3ex}
  \end{array}\eqno(B.7)$$
With these relations and the following set of fusion numbers it is
straight forward to calculate all the $N_{\alpha\beta}^{\gamma}$.
All sums of indices in the Kronecker symbols are understood to be taken
modulo $2k$ for positive indices or modulo $2k+2$ for negative ones. Here
always $j,j',j''\in\{1,\ldots,k-1\}$ and $-j,-j',-j''\in\{-1,\ldots,-k\}$.
Also note that we only distinguish the degenerate representations by
the usage of $E$. Thus, normally the choice of one of the degenerate
representations is arbitrary, only if two or three indices belong to
degenerate representations the hat ($\widehat{\phantom{x}}$) indicates,
where the relatively other choice has to be made.
Finally, we write $N(\alpha,\beta;\gamma)$ instead of
$N_{\alpha\beta}^{\gamma}$ for better readibility.\par
{\footnotesize
  $$\begin{array}{lcl}
  N(\alpha,\beta;0)           & = & \delta_{\alpha+\bar{\beta},0}
    \ =\ C_{\alpha\beta} \strut{-1.5ex}{1.5ex}\\
  N(j,j';j'')                 & = & 2 + \delta_{j+j'-j'',0}
    + \delta_{j'+j''-j,0} + \delta_{j''+j-j',0} + \delta_{j+j'+j'',0}
    \strut{-1.5ex}{1.5ex}\\
  N(j,k;j'')                  & = & \frac{1}{2}(2 + \delta_{j+k-j'',0}
    + \delta_{k+j''-j,0} + \delta_{j''+j-k,0} + \delta_{j+k+j'',0})
    \strut{-1.5ex}{1.5ex}\\
  N(j,k+1;j'')                & = & 2 + \delta_{j-j'',0}
    \strut{-1.5ex}{1.5ex}\\
  N(j,-j';j'')                & = & 2 \strut{-1.5ex}{1.5ex}\\
  N(j,-k-1;j'')               & = & 1 \strut{-1.5ex}{1.5ex}\\
  N(k,k;j'')                  & = & \frac{1}{2}(1 + (-1)^{j''+k})
    \strut{-1.5ex}{1.5ex}\\
  N(k,\widehat{k};j'')        & = & \frac{1}{2}(1 - (-1)^{j''+k})
    \strut{-1.5ex}{1.5ex}\\
  N(k,k+1;j'')                & = & 1 \strut{-1.5ex}{1.5ex}\\
  N(k,-j';j'')                & = & 1 \strut{-1.5ex}{1.5ex}\\
  N(k,-k-1;j'')               & = & \frac{1}{2}(1 + (-1)^{j''})
    \strut{-1.5ex}{1.5ex}\\
  N(k,\widehat{-k-1};j'')     & = & \frac{1}{2}(1 - (-1)^{j''})
    \strut{-1.5ex}{1.5ex}\\
  N(k+1,k+1;j'')              & = & 2 \strut{-1.5ex}{1.5ex}\\
  N(k+1,-j';j'')              & = & 2 \strut{-1.5ex}{1.5ex}\\
  N(k+1,-k-1;j'')             & = & 1 \strut{-1.5ex}{1.5ex}\\
  N(-j,-j';j'')               & = & 2 \strut{-1.5ex}{1.5ex}\\
  N(-j,-k-1,+;j'')            & = & 1 \strut{-1.5ex}{1.5ex}\\
  N(-k-1,-k-1;j'')            & = & \frac{1}{2}(1 + (-1)^{j''+k})
    \strut{-1.5ex}{1.5ex}\\
  N(-k-1,\widehat{-k-1};j'')  & = & \frac{1}{2}(1 - (-1)^{j''+k})
    \strut{-1.5ex}{1.5ex}\\
  N(k,k;k)                    & = & \frac{1}{2}(1 + (-1)^{k}
    \strut{-1.5ex}{1.5ex}\\
  N(k,\widehat{k};k)          & = & 0 \strut{-1.5ex}{1.5ex}\\
  N(k,k+1;k)                  & = & 1 \strut{-1.5ex}{1.5ex}\\
  N(k,-j';k)                  & = & \frac{1}{2}(1 - (-1)^{j'})
    \strut{-1.5ex}{1.5ex}\\
  N(k,-k-1;k)                 & = & \frac{1}{2}(1 + (-1)^{k})
    \strut{-1.5ex}{1.5ex}\\
  N(k,\widehat{-k-1};k)       & = & 0 \strut{-1.5ex}{1.5ex}\\
  N(\widehat{k},\widehat{k};k)& = & \frac{1}{2}(1 - (-1)^{k})
    \strut{-1.5ex}{1.5ex}\\
  N(\widehat{k},k+1;k)        & = & 1 \strut{-1.5ex}{1.5ex}\\
  N(\widehat{k},-j';k)        & = & \frac{1}{2}(1 + (-1)^{j'})
    \strut{-1.5ex}{1.5ex}\\
  N(\widehat{k},-k-1;k)       & = & \frac{1}{2}(1 - (-1)^{k})
    \strut{-1.5ex}{1.5ex}\\
  N(\widehat{k},\widehat{-k-1};k) & = & 0 \strut{-1.5ex}{1.5ex}\\
  N(k+1,k+1;k)                & = & 0 \strut{-1.5ex}{1.5ex}\\
  N(k+1,-j';k)                & = & 1 \strut{-1.5ex}{1.5ex}\\
  N(k+1,-k-1;k)               & = & \frac{1}{2}(1 + (-1)^{k})
    \strut{-1.5ex}{1.5ex}\\
  N(k+1,\widehat{-k-1};k)     & = & \frac{1}{2}(1 - (-1)^{k})
    \strut{-1.5ex}{1.5ex}\\
  N(-j,-j';k)                 & = & 1 \strut{-1.5ex}{1.5ex}\\
  N(-j,-k-1,+;k)              & = & \frac{1}{2}(1 + (-1)^{j+k})
  \end{array}$$ 
  $$\begin{array}{lcl}
  N(-j,\widehat{-k-1};k)      & = & \frac{1}{2}(1 - (-1)^{j+k})
    \strut{-1.5ex}{1.5ex}\\
  N(-k-1,-k-1;k)              & = & 0 \strut{-1.5ex}{1.5ex}\\
  N(-k-1,\widehat{-k-1};k)    & = & 0 \strut{-1.5ex}{1.5ex}\\
  N(\widehat{-k-1},\widehat{-k-1};k) & = & 1 \strut{-1.5ex}{1.5ex}\\
  N(k+1,k+1;k+1)              & = & 2 \strut{-1.5ex}{1.5ex}\\
  N(k+1,-j';k+1)              & = & 2 \strut{-1.5ex}{1.5ex}\\
  N(k+1,-k-1;k+1)             & = & 1 \strut{-1.5ex}{1.5ex}\\
  N(-j,-j';k+1)               & = & 2 - \delta_{j-j',0}
    \strut{-1.5ex}{1.5ex}\\
  N(-j,-k-1;k+1)              & = & 1 \strut{-1.5ex}{1.5ex}\\
  N(-k-1,-k-1;k+1)            & = & 0 \strut{-1.5ex}{1.5ex}\\
  N(-k-1,\widehat{-k-1};k+1)  & = & 0   \strut{-1.5ex}{1.5ex}\\
  N(-j,-j';-j'')              & = & 2 - \delta_{j+j'-j'',0}
    - \delta_{j'+j''-j,0} - \delta_{j''+j-j',0} - \delta_{j+j'+j'',0}
    \strut{-1.5ex}{1.5ex}\\
  N(-j,-k-1;-j'')             & = & \frac{1}{2}(2 - \delta_{j+(k+1)-j'',0}
    - \delta_{(k+1)+j''-j,0} - \delta_{j''+(k+1)-j',0}
    - \delta_{j+(k+1)+j'',0})
    \strut{-1.5ex}{1.5ex}\\
  N(-k-1,-k-1;-j'')           & = & \frac{1}{2}(1 - (-1)^{j''+k})
    \strut{-1.5ex}{1.5ex}\\
  N(-k-1,\widehat{-k-1};-j'') & = & \frac{1}{2}(1 + (-1)^{j''+k})
    \strut{-1.5ex}{1.5ex}\\
  N(-k-1,-k-1;-k-1)           & = & \frac{1}{2}(1 + (-1)^{k})
    \strut{-1.5ex}{1.5ex}\\
  N(-k-1,\widehat{-k-1};-k-1) & = & 0 \strut{-1.5ex}{1.5ex}\\
  N(\widehat{-k-1},\widehat{-k-1};-k-1) & = &
    \frac{1}{2}(1 - (-1)^{k})
  \end{array}$$
  }{\raggedleft(B.8)\par}\sn      
The automorphisms of the fusion rules can be read off from the
decompositions of $k$ into two coprime factors and similarily for $k+1$.
Without loss of generality let us assume that $k = pq$ with $p,q$ coprime.
Then we have an automorphism of the fusion algebra, namely
$j\mapsto pj$ mod $2q$, $j \in \{0,1,\ldots,k\}$ and all other labels
are left unchanged (in particular the label $k+1$ has to be considered as
zero). In particular
$N_{pj,pj'}^{pj''} = N_{j,j'}^{j''}$ where all indices are taken modulo
$2q$ and $j,j',j'' \in \{1,\ldots,k-1\}$, as can be seen directly from
the explicit form (B.8) of these fusion numbers.
If $k+1$ has such a decomposition, $k+1 = pq$, then there is an automorphism
$-j \mapsto -pj$ mod $2q$, $-j \in \{0,-1,\ldots,-k-1\}$
and again all other labels have to be left unchanged.\par
  \sn The one-one correspondence of theories and automorphisms, together
with our arguments of section 5, assure that there are no other non-trivial
automorphisms. In fact, a theory with $2R_1^2 = \frac{P_1}{Q_1}$ and
$2R_2^2 = \frac{P_2}{Q_2}$ in partition function (5.1) such that
$P_{2}Q_{2} - P_{1}Q_{1} = 1$ yields an automorphism, as described above,
and the set of these theories is complete.
  \bn\bn\bn
  \smallkapitel{{\boldmath [\ ]}}{References}{}\noindent
  \def\qq#1{\item[\lb#1\rb]}
  \def\qi{\item}
  \newcounter{rf}
  \setcounter{rf}{1}
{\small
\setbox7=\hbox{[99]\hspace{1cm}}
\begin{list}{[\arabic{rf}]}{\usecounter{rf}
  \setlength{\leftmargin}{\wd7}\setlength{\parsep}{0pt}
  \setlength{\itemsep}{0.7ex}\setlength{\labelsep}{1cm}}
\qq{1}  Alvarez-Gaum\'{e}, L., Zaugg, Ph.,
        {\em Structure Constants in the $N=1$ Superoperator Algebra},
        preprint CERN-TH.6242/91, UGVA-PHY-09-745/91 (1991)
\qq{2}  Anderson, G., Moore, G.,
        {\em Rationality in Conformal Field Theory},
        Commun.\ Math.\ Phys.\ {\bf 117} 441-450 (1988)
\qq{3}  Belavin, A.A., Polyakov, A.M., Zamolodchikov, A.B.,
        {\em Infinite Conformal Symmetry in Two-Dimensional Quantum Field
        Theory},
        Nucl.\ Phys.\ {\bf B241} 333-380 (1984)
\qq{4}  Blumenhagen, R., Eholzer, W., Honecker, A., H\"ubel, R.,
        {\em New $N=1$ Extended Superconformal Algebras with Two and Three
        Generators},
        preprint BONN-HE-92-02 (1992)
\qq{5}  Blumenhagen, R., Flohr, M., Kliem, A., Nahm, W., Recknagel, A.,
        Varnhagen, R.,
        {\em $\w$-Algebras with Two and Three Generators},
        Nucl.\ Phys.\ {\bf B361} 255-289 (1991)
\qq{6}  Bouwknegt, P.,
        {\em Extended Conformal Algebras},
        Phys.\ Lett.\ {\bf B207} 295-298 (1988)
\qq{7}  Cappelli, A.,
        {\em Modular Invariant Partition Functions of Superconformal Theories},
        Phys.\ Lett.\ {\bf B185} 82-88 (1987)
\qq{8}  Cappelli, A., Itzykson, C., Zuber, J.-B.,
        {\em Modular Invariant Partition Functions in two Dimensions},
        Nucl.\ Phys.\ {\bf B280}[FS18] 445-465 (1987),
        {\em The A-D-E Classification of Minimal and $A_1^{(1)}$ Conformal
        Invariant Theories},
        Commun.\ Math.\ Phys.\ {\bf 113} 1-26 (1987)
\qq{9}  Dijkgraaf, R., Verlinde, E., Verlinde, H.,
        {\em Conformal Field Theories on Riemann Surfaces},
        Commun.\ Math.\ Phys.\ {\bf 155} 649-690 (1988)
\qq{10} Dolan, L., Goddard, P., Montague, P.,
        {\em Conformal Field Theory of Twisted Vertex Operators},
        Nucl.\ Phys.\ B{\bf 338} 529-601 (1990),
        {\em Conformal Field Theory, Triality and the Monster Group},
        Phys.\ Lett.\ {\bf B236} 165-172 (1990)
\qq{11} Dotsenko, Vl.S.,
        {\em Three-Point Correlation Functions of the Minimal Conformal
        Theories Coupled to 2d Gravity},
        Mod.\ Phys.\ Lett.\ {\bf A6} 3601-3612 (1991)
\qq{12} Dotsenko, Vl.S., Fateev, V.A.,
        {\em Conformal Algebra and Multipoint Correlation Functions in 2d
        Statistical Models},
        Nucl.\ Phys.\ {\bf B249}[FS12] 312-348 (1984),
        {\em Four-Point Correlation Functions and the Operator Algebra in 2d
        Conformal Invariant Theories with Central Charge $c \leq 1$},
        Nucl.\ Phys.\ {\bf B251}[FS13] 691-734 (1985),
        {\em Operator Algebra of Two-Dimensional Conformal Theories with
        Central Charge $c \leq 1$},
        Phys.\ Lett.\ {\bf B154} 291-295 (1985)
\qq{13} Eholzer, W., Flohr, M., Honecker, A., H\"ubel, R., Nahm, W.,
        Varnhagen, R.,
        {\em Representations of $\w$-Algebras with Two Generators and New
        Rational Models},
        preprint BONN-HE-91-22 (1991)
\qq{14} Eholzer, W., Honecker, A., H\"ubel, R.,
        {\em Representations of $N=1$ Extended Superconformal Algebras},
        in preparation
\qq{15} Feigin, B.L., Fuks, D.B.,
        {\em Invariant Skew-Symmetric Differential Operators on the Line and
        Verma Modules over the Virasoro Algebra},
        Funkt.\ Anal.\ Appl.\ {\bf 16} 114-126 (1982),
        {\em Verma Modules over the Virasoro Algebra},
        Funct.\ Anal.\ \& Appl.\ {\bf 17} 241-242 (1983),
        {\em Verma Modules over the Virasoro Algebra},
        in: Topology, Proceedings, Leningrad 1982.\
        Faddeev, L.D., Mal'cev, A.A., (eds.),
        Lecture Notes in Mathematics, vol.\ {\bf 1060} (1984) 230.\
        Berlin, Heidelberg, New York: Springer 1984
\qq{16} Felder, G.,
        {\em BRST Approach to Minimal Models},
        Nucl.\ Phys.\ {\bf B317} 215-237 (1989), {\em Erratum},
        Nucl.\ Phys.\ {\bf B324} 548 (1989)
\qq{17} Felder, G., Fr\"ohlich, J., Keller, G.,
        {\em Braid Matrices and Structure Constants for Minimal Conformal
        Models},
        Commun.\ Math.\ Phys.\ {\bf 124} 647-664 (1989)
\qq{18} Figueroa-O'Farrill, J.M., Schrans, S.,
        {\em The Spin 6 Extended Conformal Algebra},
        Phys.\ Lett.\ {\bf B245} 471-476 (1990)
\qq{19} Figueroa-O'Farrill, J.M., Schrans, S.,
        {\em The Conformal Bootstrap and Super $\w$-Algebras},
        Int.\ J.\ Mod.\ Phys.\ {\bf A7} 591-618 (1992)
\qq{20} Ginsparg, P.,
        {\em Curiosities at $c = 1$},
        Nucl.\ Phys.\ {\bf B295}[FS21] 153-170 (1988)
\qq{21} Goulian, M., Li, M.,
        {\em Correlation Functions in Liouville Theory},
        Phys.\ Rev.\ Lett.\ {\bf 66} 2051-2055 (1991)
\qq{22} Hamada, K., Takao, M.,
        {\em Spin 4 Current Algebra},
        Phys.\ Lett.\ {\bf B209} 247-251 (1988), {\em Erratum},
        Phys.\ Lett.\ {\bf B213} 564 (1988)
\qq{23} Hornfeck, K., Ragoucy, E.,
        {\em A Coset Construction for the Super-$\w_3$ Algebra},
        Nucl.\ Phys.\ {\bf B340} 225-244 (1990)
\qq{24} Inami, T., Matsuo, Y., Yamanaka, I.,
        {\em Extended Conformal Algebras with $N = 1$ Supersymmetry},
        Phys.\ Lett.\ {\bf B215} 701-705 (1988)
\qq{25} Kausch, H.G.,
        {\em Extended Conformal Algebras Generated by a Multiplet of
        Primary Fields},
        Phys.\ Lett.\ {\bf B259} 448-455 (1991)
\qq{26} Kausch, H.G., Watts, G.M.T.,
        {\em A Study of $\w$-Algebras using Jacobi Identities},
        Nucl.\ Phys.\ {\bf B354} 740-768 (1991)
\qq{27} Kiritsis, E.B.,
        {\em Proof of the Completeness of the Classification of Rational
        Conformal Theories with $c=1$},
        Phys.\ Lett.\ {\bf B217} 427-430 (1989),
        {\em Some Proofs on the Classification of Rational Conformal Field
        Theories with $c=1$},
        preprint CALT-68-1510 (1988)
\qq{28} Kitazawa, Y., Ishibashi, N., Kato, A., Kobayashi, K., Matsuo, Y.,
        Odake, S.,
        {\em Operator Product Expansion Coefficients in $N = 1$ Superconformal
        Theory and Slightly Relevant Pertubation},
        Nucl.\ Phys.\ {\bf B306} 425-444 (1988)
\qq{29} Kliem, A.,
        {\em Konstruktion von $\w$-Algebren},
        Diplomarbeit BONN-IR-91-46 (1991)
\qq{30} Komata, S., Mohri, K., Nohara, H.,
        {\em Classical and Quantum Extended Superconformal Algebra},
        Nucl.\ Phys.\ {\bf B359} 168-200 (1991)
\qq{31} Kontsevitch, M.,
        {\em Rational Conformal Field Theory and Invariants of 3-Dimensional
        Ma\-ni\-folds},
        preprint CPT-88/P.2189 (1988)
\qq{32} Mathur, S.D., Mukhi, S., Sen, A.,
        {\em Reconstruction of Conformal Field Theories from Modular Geometry
        on the Torus},
        Nucl.\ Phys.\ {\bf B318} 483-540 (1989)
\qq{33} Moore, G., Seiberg, N.,
        {\em Polynomial Equations for Rational Conformal Field Theories},
        Phys.\ Lett.\ {\bf B212} 451-460 (1988)
\qq{34} Nahm, W.,
        {\em Chiral Algebras of Two-Dimensional Chiral Field Theories
        and their Nornal Ordered Products}, in:
        Recent Developments in Conformal Field Theories, Proceedings,
        ICTP, Trieste, 1989. Randjbar-Daemi, S., Sezgin, E., Zuber, J.-B.,
        (eds.). Singapore, New Jersey, London, Hong Kong: World Scientific
1990,
        {\em Conformal Quantum Field Theories in Two Dimensions},
        World Scientific, to be published
\qq{35} Schellekens, A.N., Yankielowicz, S.,
        {\em Extended Chiral Algebras and Modular Invariant Partition
        Functions},
        Nucl.\ Phys.\ {\bf B327} 673-703 (1989)
\qq{36} Varnhagen, R.,
        {\em Characters and Representations of New Fermionic $\w$-Algebras},
        Phys.\ Lett.\ {\bf B275} 87-92 (1992)
\qq{37} Verlinde, E.,
        {\em Fusion Rules and Modular Transformations in 2d Conformal Field
        Theory},
        Nucl.\ Phys.\ {\bf B300}[FS22] 360-376 (1988)
\qq{38} Witten, E.,
        {\em Quantum Field Theory and the Jones Polynomial},
        Commun.\ Math.\ Phys.\ {\bf 122} 351-399 (1989)
\qq{39} Zamolodchikov, Al.B.,
        {\em Infinite Additional Symmetries in Two-Dimensional
        Conformal\linebreak Quantum Field Theory},
        Theor.\ Math.\ Phys.\ {\bf 65} 1205-1213 (1985)
\qq{40} Zhang, D.-H.,
        {\em Spin-4 Extended Conformal Algebra},
        Phys.\ Lett.\ {\bf B232} 323-326 (1989)
\end{list}}
\end{document}